\documentclass[useAMS]{mn2e}

\usepackage{graphicx}
\usepackage{amssymb}

\title{Simulating Wide-Field Quasar Surveys from the Optical to Near-Infrared}

\author[N. Maddox \& P. Hewett]{Natasha Maddox\thanks{nmaddox@ast.cam.ac.uk},
  Paul C. Hewett
\vspace*{6pt}\\
Institute of Astronomy, University of Cambridge, Madingley Road,
Cambridge CB3 0HA, UK \\}

\begin{document}


\maketitle

\begin{abstract}

A number of deep, wide-field, near-infrared surveys employing new
infrared cameras on 4m-class telescopes are about to commence. These
surveys have the potential to determine the fraction of luminous
dust-obscured quasars that may have eluded surveys undertaken at
optical wavelengths. In order to understand the new observations it is
essential to make accurate predictions of surface densities and
number-redshift relations for unobscured quasars in the near-infrared
based on information from surveys at shorter wavelengths. The accuracy
of the predictions depends critically on a number of key
components. The commonly used single power-law representation for
quasar SEDs is inadequate and the use of an SED incorporating the
upturn in continuum flux at $\lambda \sim 12\,000\,$\AA \ is
essential. The presence of quasar host galaxies is particularly
important over the restframe wavelength interval $8000 < \lambda <
16\,000\,$\AA \ and we provide an empirical determination of the
magnitude distribution of host galaxies using a low redshift sample of
quasars from the SDSS DR3 quasar catalogue. A range of models for the
dependence of host galaxy luminosity on quasar luminosity is
investigated, along with the implications for the near-infrared
surveys. Even adopting a conservative model for the behaviour of host
galaxy luminosity the number counts for shallow surveys in the $K$
band increase by a factor of two. The degree of morphological
selection applied to define candidate quasar samples in the
near-infrared is found to be an important factor in determining the
fraction of the quasar population included in such samples.

\end{abstract}



\section{Introduction}

There has been much debate in recent years regarding the existence, or
otherwise, of a significant population of dust-obscured quasars that
have eluded surveys based on an ultraviolet excess or
optical colours. Since gas is required to fuel quasars and given the
success of unified models, which invoke non-spherical structures that
will certainly obscure the line-of-sight to the inner regions of
certain active galactic nuclei (AGN), the existence of an obscured
population of quasars is a natural consequence of our present
understanding of AGN and quasars. Indeed, examples of obscured objects
have been identified, from weakly reddened quasars, whose spectra show
small deviations consistent with obscuration by dust at ultraviolet
wavelengths (e.g. Richards et al. 2003), to more heavily obscured
objects where emission at ultraviolet wavelengths is almost completely
absent (e.g. Gregg et al. 2002). Since these objects have been
definitively shown to exist, the debate turns to the question of their
numbers in relation to the total quasar population.

Follow-up observations of the source populations identified from the
very deep {\it Chandra} X-ray fields (e.g. Alexander et al. 2003) have
established that a large population of optically obscured AGN at
bolometric luminosities $L_{Bol} \la 10^{45}\,{\rm ergs}\,{\rm
s}^{-1}$ exist at redshifts $z \la 2$ (Barger et al. 2005). At higher
luminosities, in the regime probed by the majority of large
optically-selected quasar surveys (e.g. Hewett, Foltz \& Chaffee 1995;
Croom et al. 2004) the surface density of quasars is low and the X-ray
catalogues contain relatively few objects, limiting the conclusions
that can be drawn. In fact, Barger et al. (2005) find no evidence that
a significant fraction of the quasar population is obscured at high
luminosities, a finding apparently at variance with the conclusions of
the widely cited study of Webster et al. (1995).

The {\it Spitzer Space Telescope} has recently enabled deep,
high-quality observations at infrared wavelengths to be performed. A
type I quasar luminosity function (QLF) for the rest-frame 8$\mu$m has
been constructed by Brown et al. (2005), spanning $1<z<5$. The study
specifically eliminated $z<1$ objects to avoid complications due to
the presence of the quasar host galaxies. The shape of the derived
infrared QLF is very similar to those constructed from
optically-selected quasar samples. Comparison between surveys reveals
that $\sim$20 per cent of quasars present in the infrared sample are
missing from optically-selected samples, but the missing fraction
increases with decreasing luminosity. Most of the objects missing from
the optical samples have red colours.

The conclusions drawn from a series of relatively modest samples of
reddened quasars vary significantly. At one extreme, it has been
suggested that the number of reddened objects far exceeds the number
of optically-selected quasars (Webster et al. 1995), or that the
obscured objects that have been found represent only `the most
luminous tip of the red quasar iceberg' (White et al. 2003). However,
other surveys (Francis, Nelson \& Cutri 2004) claim that reddened
quasars are not the dominant population. Targeted searches for type II
quasars, the high-luminosity equivalents of the type II Seyfert
population, predicted to exist in a number of unified-scheme models,
have also begun to produce results (Zakamska et al. 2003, 2005) but
the number of objects remains small.

The ability to undertake effective surveys for luminous quasars that
are subject to a significant degree of reddening is about to be
enhanced dramatically through the availability of large-format
infrared cameras on 4m-class telescopes. In particular, the new Wide
Field Camera (WFCAM) on the United Kingdom Infrared Telescope (UKIRT)
on Mauna Kea has been commissioned successfully and the first
observations for a series of photometric surveys under the banner of
the UKIRT Infrared Deep Sky Survey (UKIDSS) have already begun. Of
particular interest is the Large Area Survey (LAS) that will cover
4000 square degrees of the Sloan Digital Sky Survey (SDSS) to a depth
more than three magnitudes fainter ($K \simeq 18.5$) than the
Two-Micron All Sky Survey (2MASS) (Cutri et al.  2003). Some 2000
square degrees of the LAS are scheduled for completion by mid-2007.
Additional major wide-field near-infrared surveys will be undertaken
by the WIRCam instrument on the Canada-France-Hawaii Telescope and the
ESO VISTA telescope at Paranal.

For largely historical reasons almost all investigations of the form
and evolution of the quasar luminosity function at redshifts $z<3$
from optically-selected surveys have been based on samples identified
at blue wavelengths, employing flux limits in $B$, or, more recently,
$b_J$ passbands. Quasar surveys in near-infrared passbands will almost
certainly reveal the existence of a population of reddened objects
that are, at best, under-represented in the traditional $B$-band
samples. However, the predicted form of the number-magnitude and
number-redshift relations for such surveys based purely on the known
properties of quasars included in the $B$-band samples is necessary in
order to interpret the new survey results. The combination of the very
steep luminosity function at bright quasar luminosities, the form of
the quasar SEDs (including strong emission lines) and the presence of
host galaxies, which become increasingly prominent at longer
wavelengths, means that the nature of the number-magnitude and
number-redshift relations at near-infrared wavelengths are not as
straightforward as they might at first seem. Certainly, the
predictions are very different from those based on the approximation
of a simple power-law SED for the quasars with a typical colour of
$B-K \simeq 2.5$ that is still commonly employed (e.g. Glikman et
al. 2004).

The primary purpose of this paper is to take our current knowledge
describing (i) the form and evolution of the quasar luminosity
function from optical surveys, (ii) the shape of the ultraviolet
through near-infrared SED of luminous unreddened quasars and (iii) the
effect of host galaxy light on the magnitudes and colours of quasars,
to produce predictions for the form of the number-magnitude and
number-redshift relations expected from imminent wide-field
near-infrared surveys. The predictions provide a
reference against which the observational results may be compared,
allowing the space density and properties of the new reddened
population of quasars to be quantified.

The outline of the paper is as follows. Section 2 describes the input
data required for the simulations, $\S$3 discusses the estimation of
quasar magnitudes from the SDSS data where the presence of a host
galaxy leads to a non-stellar classification for the object in SDSS,
and $\S$4 briefly outlines some of the details of the simulation
program. The results are presented in $\S$5, with discussion and
conclusions following in \S6 and \S7, respectively. A determination of
the $b_J$ passband is described in Appendix A. Appendix B includes
tabular material giving the predicted number-magnitude counts for a
number of passbands. Concordance cosmology with $H_{0} = 70$ km
s$^{-1}$ Mpc$^{-1}$, $\Omega_{m} = 0.3$ and $\Omega_{\Lambda} = 0.7$
is assumed throughout. Magnitudes on the Vega system are used
throughout the paper, with the SDSS AB-magnitudes converted to the
Vega system using the relations: $u=u_{\rm AB}-0.93$, $g=g_{\rm
  AB}+0.10$, $r=r_{\rm AB}-0.15$, $i=i_{\rm AB}-0.37$, and $z=z_{\rm
  AB}-0.53$.


\section{Setting up the Simulations}

In addition to a quasar luminosity function, the simulations
require a number of other inputs, including the evolution of the QLF
with redshift, the evolution of the quasar SED as a function of
redshift and changes to the luminosity and colour of the quasars due
to the presence of host galaxies. Predictions for specific surveys are
made by incorporating appropriate errors in the measured magnitudes
and applying object selection as a function of apparent magnitude,
absolute magnitude, redshift and object morphology. Predictions for
any specified passband can be made by performing transformations
between passbands using detailed specifications of individual
passbands, including filter transmission, detector quantum efficiency
and atmospheric transmission. References to the source of the SDSS and
2MASS passbands and the procedures used to calculate the UKIDSS
passbands used in the paper are given in Hewett et al. (2006). The
2MASS passband employed here is the 2MASS-specific version of the
$K_s$ passband, and is denoted $K_{\rm 2MASS}$. In the following
subsections the principal inputs to the simulation are described.

\subsection{Quasar Luminosity Function}

The shape of the QLF, the values of its parameters and its evolution
up to redshift $z\approx$ 2 were taken from the 2dF and 6dF QSO
Redshift Surveys (2QZ+6QZ; Croom et al., 2004). These two surveys
contain more than 20\,000 quasars, selected from APM scans of $ub_Jr$
UK Schmidt Telescope photographic plates as blue point sources,
spanning $16 < m_{b_J} < 20.85$ and a range in redshift of $0 < z
\lesssim 3$. Data used to construct the QLF is restricted to $0.4 < z
< 2.1$, including nearly 16\,000 objects. The relatively high lower
redshift limit reduces contamination from intrinsically faint quasars
where light from the host galaxy may artificially increase the
apparent brightness, causing objects to move above the faint magnitude
limit. The luminosity function is parametrized as a double power-law
in magnitude, with bright and faint slopes of $\alpha=-3.25$ and
$\beta=-1.01$, respectively, a characteristic break magnitude
$M^{*}_{b_J}=-20.47$, and normalisation density at redshift zero
$\Phi(M^{*}_{b_J})=1.84\times 10^{-6}$:

\[
\Phi(M_{b_{J}},z) = \frac{\Phi(M^{*}_{b_{J}})}{10^{0.4(\alpha+1)(M_{b_{J}}-M^{*}_{b_{J}})}+10^{0.4(\beta+1)(M_{b_{J}}-M^{*}_{b_{J}})}}
\]

Note that the numerator is actually twice the value of the space
density at $M^{*}_{b_J}$. The evolution of the QLF is divided into
three redshift regimes, with $M^{*}$ evolving exponentially in
brightness with redshift for $0.0 < z \le 2.2$, no evolution from $2.2
< z \le 3.0$, then $\Phi(M^{*})$ declining exponentially as given in
Fan et al. (2001) for redshifts $z > 3$. The Croom et al. (2004) QLF
is consistent with much earlier work and both the QLF-shape and form
of the evolution to $z \approx 2$ are well established, providing a
well-determined reference for the unobscured quasar population.

\subsection{Quasar Spectral Energy Distributions}\label{SED}

The completion of relatively large surveys for quasars at optical
wavelengths, such as the Large Bright Quasar Survey (LBQS) and FIRST
Bright Quasar Survey (FBQS) that involve the acquisition of fluxed
spectra of at least intermediate signal-to-noise ratio (Hewett et
al. 1995; White et al. 2000) has allowed the construction of high
signal-to-noise ratio composite quasar spectra (Francis et al. 1991;
Brotherton et al. 2001). The SDSS has taken the subject area to a new
level, increasing both the number of quasars (Schneider et al. 2005)
and quality of the spectra (Stoughton et al. 2002; Abazajian et
al. 2003) dramatically. Investigating the properties and incidence of
quasars suffering from the effects of attenuation by dust (Richards et
al. 2003) is also possible. However, to generate baseline predictions
based on the properties of quasars detected at optical wavelengths, it
is the generic composite spectra, dominated by the characteristics of
the typical quasars included in the surveys, that are of interest.

The ultraviolet properties of the composite quasar SEDs from the LBQS,
FBQS and SDSS are in close agreement (Vanden Berk et al. 2001). The
optical portion of the composite quasar SEDs is complicated by the
presence of significant host galaxy contributions in the relatively
low-luminosity quasars at low redshifts ($z \le 0.4$) that dominate
the composite spectra longward of $\sim 5000\,$\AA. Even if the quasar
and host galaxy contributions can be disentangled the red wavelength
limit of the SDSS-composite spectra reaches only $\sim
8500\,$\AA. Information on the form of the optically-selected quasar
SED to much longer wavelengths is necessary in order to provide
predictions for surveys undertaken at near-infrared wavelengths.

2MASS (Cutri et al. 2003) provides such information for the LBQS and
FBQS surveys as well as bright subsets of the SDSS quasars. However,
the large epoch differences between 2MASS and the LBQS and FBQS means
a bright subsample of the SDSS DR3 quasar catalogue (Schneider et al.
2005) provides the most stringent constraints on the overall form of
the restframe ultraviolet to near-infrared SED of bright
optically-selected quasars.

Constraints on the form of the quasar SED used in the simulations
presented here were obtained by utilising the bright quasars in the
SDSS DR3 quasar catalogue, a large fraction of which also possess
$JHK$ photometry from 2MASS. The 2MASS flux limit is relatively bright
and the fraction of SDSS quasars that possess 2MASS detections
decreases monotonically with increasing SDSS $i$-band magnitude.
Starting with a bright magnitude limit of $i=14.5$, a sample of SDSS
quasars was selected by requiring that 90 per cent of the sample have
corresponding 2MASS magnitudes. The sample was restricted in redshift
to the range $0.1 \le z \le 3.6$. The low redshift limit was adopted
in order to reduce potential contamination from host galaxies. The
number of bright quasars with $z > 3.5$ is small and the actual value
adopted for the high redshift limit has essentially no effect on the
conclusions. The resulting sample consists of 3022 objects, magnitudes
$14.6 \le i \le 17.39$, redshifts $0.1 \le z \le 3.6$, of which (by
definition) 90 per cent also possess 2MASS $JHK$ magnitudes. Among the
sample, 314 quasars, almost all with redshifts $z < 0.4$, are
classified as non-stellar according to the SDSS photometry.

A relatively simple parametric model is adopted to describe the quasar
SED, the main components of which are: i) an underlying power-law
continuum, ii) an emission line spectrum, and iii) an optically thick
Balmer continuum emission component. The individual parameters were
systematically varied until a good fit to the median colours $u-g$,
$g-r$,..., $H-K$ of the SDSS--2MASS quasar sample was obtained. The
median colours are not affected by the small fraction of objects that
do not possess 2MASS $JHK$ magnitudes as the appropriate limits to the
colours of the quasars are incorporated into the calculation of the
median colours. The resulting continuum is represented by two
power-laws of the form $F(\nu) \propto \nu^\alpha$, with $\alpha=-0.3$
from $912<\lambda<12\,000$\,\AA\, and $\alpha=-2.4$ from
$12\,000<\lambda<25\,000$\,\AA. The emission line spectrum, which
incorporates optical and ultraviolet emission from Fe{\tt II}
multiplets, is taken from the LBQS composite of Francis et al. (1991)
extended to $7000\,$\AA \, to include the H$\alpha$ emission line. In
addition, a Pa$\alpha$ emission line is included.  The equivalent
width (EW) of the H$\alpha$ line is 380\,\AA \ for $z_{quasar} \le
0.3$ (see below). The Balmer continuum emission, in the form of
equation 6 of Grandi (1982), with the temperature set at 12\,000\,K
and an optical depth $\tau$ of 45, produces enhanced flux over the
underlying power-law continuum for $1500<\lambda<3500$\,\AA.  The
fraction of the underlying continuum at 3000\,\AA\, is 0.1. The effect
of intergalactic absorption, including Ly$\alpha$, Ly$\beta$, and
Ly$\gamma$ components, is modelled using the tabulation of optical
depths as a function of redshift given by Songaila (2004), together
with a Lyman-limit system at the quasar redshift. The only passbands
affected by the presence of intergalactic absorption are the $u$ and
$b_J$ bands for quasars with redshifts $z > 2$.

It is not obvious that such a simple parametric model, with no
dependence on the redshift of the quasars, is capable of reproducing
the colours over such an extended redshift (and luminosity) range.
However, a number of investigations (e.g. Kuhn et al. 2001; Pentericci
et al. 2003; Fan et al. 2004) have noted the lack of evidence for any
significant evolution in the mean properties of the ultraviolet and
optical SEDs of optically-selected quasars as a function of redshift.
The same result is evident from the distribution of $u$ through $K$
colours for the bright SDSS sample in that a single quasar SED
accurately reproduces the trend in the median colour of the sample
over the entire redshift range $0.1 \le z \le 3.6$.

\begin{figure}
  \resizebox{\hsize}{!}{\includegraphics{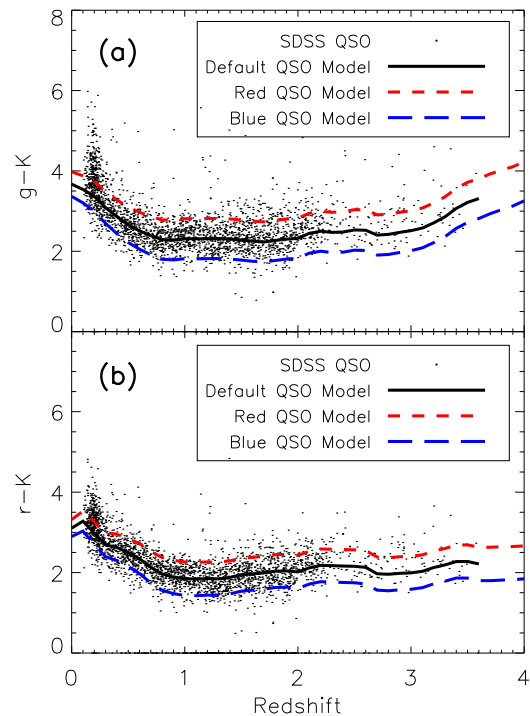}}
  \caption{Colours of the bright SDSS DR3 quasar subsample, the median
    quasar model, a red quasar model, and a blue quasar model. (a)
    $g-K$; (b) $r-K$. The cluster of redder objects at very low redshift
    are mostly flagged as `extended' in the DR3 quasar catalogue.}
  \label{fig:SDSScolours}
\end{figure}

The sole exception to this `no SED evolution' finding relates to the
strength of the H$\alpha$ emission line. The flux-limited SDSS quasar
sample includes quasars with very different luminosities as a function
of redshift; quasars at $z \sim 2$ are a factor of $\sim$100 brighter
than the quasars at $z \sim 0.2$. The equivalent width of the
H$\alpha$ emission line from the LBQS composite spectrum applies to
quasars of low luminosity and the well-established Baldwin effect
(Baldwin 1977; Yip et al. 2004) would suggest a significant reduction
in the equivalent width of the H$\alpha$ emission line for the high
luminosity quasars at redshifts $z > 2$ is expected. The entry of the
H$\alpha$ emission line into the $K$-band at $z\approx 2.2$ provides a
direct measure of its strength at high redshift confirming the
existence of a systematic reduction in the H$\alpha$ equivalent width
with luminosity. Including a luminosity dependence\footnote{The
  luminosity dependence is parametrized using the quasar luminosity,
  corresponding to the survey magnitude limit, as a function of
  redshift. The majority of quasars are found within $\sim 1\,$mag of
  the survey flux limit and the restricted range of luminosity probed
  at a given redshift means the approximation is adequate for our
  purposes} of the H$\alpha$ emission line equivalent width of the
form ${\rm EW(H\alpha)} \propto L_{quasar}^{-0.2}$, for quasars with
redshifts $z > 0.3$, provides an excellent fit to the strength of the
H$\alpha$ line in the median quasar spectrum as the feature moves
through the $iJHK$ bands with increasing redshift (and hence quasar
luminosity). Fig. \ref{fig:SDSScolours} illustrates the locus of the
adopted quasar SED as a function of redshift for two different optical
to near-infrared (optical--NIR) colours. Also included in the figure
are the colours for the same quasar SED, except the slope of the
underlying continuum power-law shortward of $\lambda$=12\,000\,\AA\,
is changed to $\alpha=-0.6$ to model a redder quasar, and $\alpha=0.0$
for a bluer quasar.

Comparison of the model quasar SED with the recently released
optical--NIR empirical composite of Glikman, Helfand \& White (2005)
shows striking similarily, with a few differences. Most notably, the
model quasar is somewhat redder at longer wavelengths, causing the
derived K-corrections to be larger, particularly at $z<1$.

Composite, quasar plus host-galaxy SEDs are constructed via a simple
sum of quasar and galaxy fluxes. The fraction of quasar and galaxy
contributing to the combined SED is specified by the ratio of the
respective fluxes over the restframe interval $4000-5000$\,\AA.

\subsection{Host Galaxy SED/Evolution}\label{hostgalaxy}

Due to the very different shapes of quasar and galaxy SEDs, light from
the host galaxy of a quasar will be most important at longer
wavelengths, particularly around $\lambda$=12\,000\,\AA. A series of
studies (see Table~\ref{tab:gals}) have found that, for bright
quasars, the host galaxies are massive, bright ellipticals. The hosts
of less luminous AGN have been found to have disc components
(e.g. Dunlop et al. 2003; Kauffmann et al. 2003) with some evidence of
recent star formation. The majority of quasars included in the
simulations presented here are of at least moderate luminosity, and
thus likely to be hosted by an early-type galaxy. However, we have
investigated the effect of different galaxy SEDs by utilising the
redshift zero template elliptical and Sb galaxies from Mannucci et
al. (2001). The use of the two SEDs should encompass the range of
galaxy types expected at the redshifts ($z \la 1$) where the presence
of the galaxy is significant. The star formation histories of
later-type galaxies can take a variety of forms and, for simplicity,
the SEDs of both galaxy types do not evolve with redshift.

The adopted relationships between the quasar and host galaxy
luminosity are described in Section \ref{qgrelation}. One of the
relationships investigated is based on the brightness distribution of
host galaxies at $z\sim 0.2$ (Section \ref{PSFMag}), which are then
allowed to brighten with increasing redshift to take account of
passive stellar evolution. The baseline evolution of the galaxy
luminosity with redshift has been taken from a Bruzual \& Charlot
(2003, hereinafter BC2003) model of a stellar population formed at
$z=5$ in a single burst with an exponential decay $\tau =1\,$Gyr. The
passive evolution of an old stellar population is appropriate for the
elliptical SED and we perform a specific test (Section
\ref{galbright}) to confirm that the results of the simulations are
not sensitive to significantly more rapid evolution in luminosity with
redshift.

In some of the simulations, a fraction of the quasars are located in
very luminous galaxies. To ensure that such galaxies do not evolve to
become brighter than any known galaxy, a single upper limit to the
luminosity achieved by any galaxy at a given redshift is adopted. The
limit at $z=0$ is taken as $M(b_{J})=-22.4$ (i.e. 2 magnitudes brighter than
the redshift zero value for $M^{*}$ of $-20.4$ from the 2dFGRS, Norberg
et al. 2002), then adjusted at each redshift to take account of the
BC2003 model luminosity evolution.

\begin{table*}
\centering
\caption{\label{tab:gals}Quasar and host galaxy properties of several
 studies investigated. Host galaxy properties resulting from the
  simulations were compared to these values where appropriate.}
\begin{minipage}{18cm}
\begin{tabular}{lccccccc} \hline
Ref.$^{a}$ & Redshift & M$_{qso}$$^{b,c,d}$ & Radio? & Selection &
M$_{gal}$ & $R_{1/2}$$^b$ & Morphology \\ 
 & & ($b_{J}$) & & & ($b_{J}$) & (kpc) & \\ \hline
Schade & $0.03-0.15$ & $-23.5<M<-13.9$ & RQ & Xray & 
$-22.6<M<-17.8$ & $0.2-18.1$ & mostly E, few Sab \\
Dunlop & $0.10-0.25$ & $-24.7<M<-20.4$ & RQ,RL & Radio, Optical 
& $-22.5<M<-20.4$ & $2.5-14.8$ & mostly E \\
Floyd & $0.297-0.43$ & $-26.7<M<-23.1$ & RL,RQ & Optical & 
$-23.1<M<-20.8$ & $1.3-15.3$ & mostly E  \\ 
Croom & $0.1-\sim 3$ & $-28.68<M<-18.43$ & -- & Optical & 
$-27.9<M<-17.6$ & -- & assumed E SED \\ \hline
\end{tabular}
$^{a}$ {\footnotesize Schade, Boyle \& Letawsky 2000; Dunlop et
  al. 2003; Floyd et al. 2004; Croom et al. 2002}\\
$^{b}$ {\footnotesize Magnitudes and scalelengths have been converted
  to concordance cosmology}\\
$^{c}$ {\footnotesize $B$-band AB magnitudes have been converted to
  Vega magnitudes using $B_{AB}-B=-0.1$}\\
$^{d}$ {\footnotesize The following offsets have been used to convert
  from the passband the individual studies were performed in to $b_J$:
  $b_J-B=-0.055$, $b_J-V=0.09$, $b_J-R=0.52$ for quasars, and
  $b_J-B=-0.15$, $b_J-V=0.783$, $b_J-R=1.395$ for an elliptical
  galaxy. The conversions for an Sb galaxy are very similar to those
  for an elliptical galaxy.}\\
\end{minipage}
\end{table*}


\section{The Distribution of Host Galaxy Magnitudes}

Although it is clear that most bright quasars are hosted by elliptical
galaxies, the observed brightnesses of the galaxies range
significantly for a given quasar luminosity (see references contained
within Table~\ref{tab:gals}). It has been shown that the mass of the
central black hole and the mass of its host galaxy bulge component are
related via the Magorrian relation (Magorrian et al. 1998), which
would suggest a linear correlation between quasar and galaxy
magnitudes, but quasar fuelling efficiency and obscuring dust would
tend to complicate the relation. Quasars of a given magnitude may
therefore be found in galaxies of a variety of magnitudes, an
observational result further reinforced by recent simulation work by
Lidz et al. (2005), among others. The distribution of host galaxy
magnitudes for a given quasar luminosity is as yet unknown. This has
prompted us to devise a method for determining the distribution using
the information contained within the SDSS DR3 quasar catalogue.

\subsection{Host Galaxy Magnitudes from SDSS Quasars}\label{PSFMag}

In contrast to the 2QZ survey, which selects only blue point sources
as candidates, the SDSS quasar selection algorithm does not
discriminate with regard to the observed morphology of targets.
Therefore, the bright subset of DR3 quasars (Section \ref{SED}) may be
divided into point source and extended objects, and two separate
composite spectra created. The extended source composite spectrum
shows strong stellar absorption lines, indicating the presence of host
galaxy light entering into the spectroscopic fibre. The EWs of the
absorption lines, when compared to the EWs from a redshift zero
elliptical galaxy template from Mannucci et al. (2001), implies that
the host galaxies contribute $\sim 50$ per cent of the total flux to
the objects. Much weaker absorption lines, corresponding to a host
galaxy contribution of 5--10 per cent, are found in the composite
created from the objects classified as point sources. Given the large
fraction of starlight evident in the quasar spectra, the question of
the impact on the SDSS PSF magnitudes, recommended for use with point
sources, arises.

The SDSS photometric catalogue provides both Petrosian and PSF
magnitudes for sources. The Petrosian magnitudes (Petrosian 1976)
measure a close to constant fraction of the total flux received from
extended objects by defining a characteristic Petrosian radius and
summing the flux within that radius. The PSF magnitude is measured by
fitting a scaled PSF model to the object. This method provides an
accurate measurement of the flux received from point-like objects such
as stars or bright quasars. All magnitudes included in the SDSS quasar
catalogues are PSF magnitudes, which will contain a certain amount of
flux from the quasar host galaxies, particularly for low redshift,
resolved objects (Schneider et al. 2003). The presence of the host
galaxy will affect both the total flux and the colours of the objects.

\begin{figure}
  \resizebox{\hsize}{!}{\includegraphics{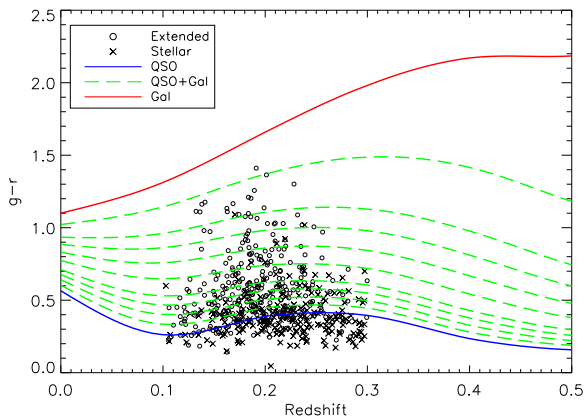}}
  \caption{PSF $g-r$ colour of the bright subsample of SDSS DR3
    quasars (crosses and circles). Also shown are the $g-r$ colours
    for an elliptical galaxy (top), the default quasar (bottom), and
    synthetic quasar+galaxy spectra for $0<R_{gq}<1$. The $R_{gq}$
    values of the dashed lines, from bottom to top, are: 0.05, 0.1,
    0.15, 0.2, 0.3, 0.4, 0.5, 0.6 and 0.8.}
  \label{fig:g-r}
\end{figure}

Fig. \ref{fig:g-r} shows the distribution of PSF $g-r$ colour {\it vs}
redshift for a low-redshift ($0.1<z<0.3$) sub-sample of the quasar
sample, defined in Section 2.2, with magnitudes $15.6 < g < 18.8$
($-24.8<M_g<-21.3$). The redshift range is narrow, so there should not
be any significant evolution effect. The solid lines show the $g-r$
colour of an elliptical galaxy (top) and a quasar created as described
in Section \ref{SED} (bottom), between $0.0<z<0.5$. The dashed lines,
from bottom to top, show the $g-r$ colours for synthetic quasars that
have increasing amounts of light from an elliptical galaxy
contributing to the overall shape of the combined SED. The relative
amount of galaxy and quasar flux is parametrized as follows, with
$R_{gq}=0$ for a quasar and $R_{gq}=1$ for a galaxy:

\[
R_{gq} = \frac{F_{galaxy}}{F_{galaxy+quasar}}
\]

where $F_{galaxy}$ is the flux of the galaxy over the restframe
interval $4000-5000\,$\AA \ and $F_{galaxy+quasar}$ is the combined
flux of the galaxy and quasar over the same wavelength interval. The
synthetic colours correspond to Petrosian measures, as they include
the total flux from both the quasar and host galaxy. Crosses mark the
objects that were classified as point sources by the SDSS photometric
pipeline (Lupton et al. 2001), while circles mark objects that were
classified as extended. The stellar sources have colours that cluster
around the $g-r$ colour calculated from the pure quasar SED, but the
non-stellar objects show a much larger spread in colour, with some
objects consistent with significant light from a host galaxy.

\begin{figure}
  \resizebox{\hsize}{!}{\includegraphics{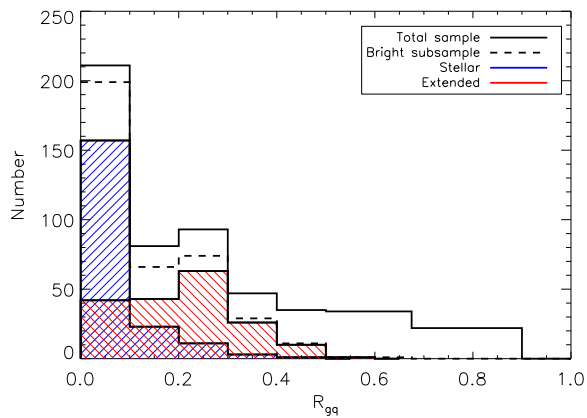}}
  \caption{Histogram of $R_{gq}$ values derived from the Petrosian
    magnitude $g-r$ colours of the bright subsample of DR3
    quasars. The $R_{gq}$ values have been binned.}
  \label{fig:Rgqhist}
\end{figure}

When the same plot is created using the Petrosian magnitudes of the
objects, the colours are skewed to larger $g-r$ values, indicating a
larger contribution to the total flux from an extended component. The
presence of i) spectral signatures due to starlight in the SDSS
spectra, ii) a systematic trend between the $g-r$ colours and the
morphological classification, iii) the strengthening of the trend when
Petrosian magnitudes are employed, provides strong support for the
presence of host galaxies as the dominant contribution to the extended
distribution of $g-r$ colours. Furthermore, the distribution of $g-r$
colours is in excellent agreement with the model predictions for the
colours of pure quasar and composite object SEDs which contain
approximately equal contributions of quasar and elliptical galaxy
light.

Fig. \ref{fig:g-r} is shown purely for illustration purposes. In order
to correctly determine the fraction of galaxy flux contributing to
each object, the Petrosian magnitudes must be used. Assuming the
object SEDs are composed of a combination of quasar and galaxy light
and that the Petrosian magnitudes provide an accurate measure of the
total flux, then the position of an object on the $(g-r)$, redshift
plane in relation to the model colours for composite SEDs gives a
measure of the relative proportions of quasar and host galaxy light
present in each object. The total object flux may then be separated
into pure galaxy and pure quasar contributions, from which the
magnitudes of each may be calculated.

As a result of the decomposition of the object magnitudes into their
component brightnesses, 30 per cent of the sample possess quasar
magnitudes that do not meet the absolute magnitude selection criterion
of $M_{i}<-22.4$ used in the compilation of the SDSS DR3 quasar
catalogue. Most of the excluded objects have large $g-r$ colours, but
a significant number lie close to the faint magnitude limit and have
$g-r$ colours consistent with only $\sim$ 30 per cent galaxy
contribution (i.e. $R_{gq}=0.3$). It will be important for
investigations of the form of the QLF at low redshifts using the SDSS
sample to incorporate the effect of host galaxies on object magnitudes.

\begin{figure}
  \resizebox{\hsize}{!}{\includegraphics{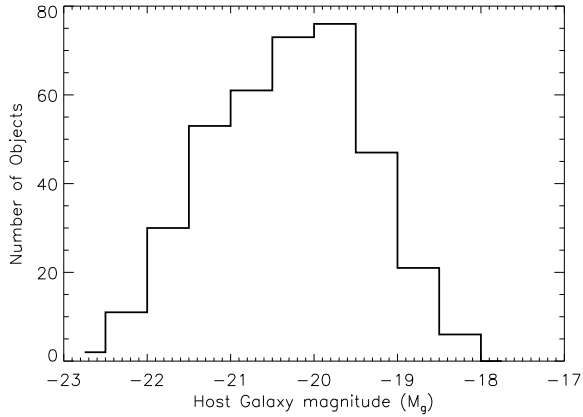}}
  \caption{Histogram of extracted host galaxy magnitudes from the bright, low
    redshift SDSS DR3 quasar sample, from the Petrosian magnitudes.}
  \label{fig:galmags}
\end{figure}

Excluding the objects that do not satisfy the absolute magnitude
criterion leaves a quasar sample, in a narrow redshift range, for
which the empirical distribution of $R_{gq}$ is now known. The
distribution of $R_{gq}$, shown in Fig. \ref{fig:Rgqhist}, is used to
determine the value and relative weighting of the constants describing
the relation between quasars and their host galaxies in Section
\ref{qgrelation}. We also have an empirical estimate, (Fig.
\ref{fig:galmags}) of the distribution of host galaxy magnitudes for
quasars found in the SDSS survey at redshift $z \sim 0.2$.

\subsection{Quasar--Host Galaxy Luminosity Relationships}\label{qgrelation}

For the simulations, each quasar must be assigned to a host galaxy of
a given brightness. Adopting a simple power law relationship, $L_{gal}
\propto L_{qso}^{\gamma}$, or equivalently, expressed in magnitudes,
$M_{gal} = A + \gamma M_{qso}$ with $A$ being a normalisation
constant, there are essentially three possible
parametrizations. First, at one extreme, the magnitudes of the quasar
and its host galaxy are unrelated, corresponding to the case of
$\gamma$=0.

But, since quasars of a given brightness are found in galaxies of a
variety of brightnesses, a distribution of initial galaxy magnitudes,
or constants $A$, is required. The empirical distribution found from
Fig. \ref{fig:galmags} is used to determine the range and relative
weighting of the constants, $A$. The SDSS $g$ passband is very similar to
the $b_{J}$ passband and $b_{J}=g$ has been adopted without the use of
a colour term. For this $\gamma$=0 case, the galaxies are then assumed
to brighten due to passive stellar evolution according to the BC2003
prescription described in Section \ref{hostgalaxy}.

Alternatively, the quasar and galaxy magnitudes may be related with $0
< \gamma \le 1$. Croom et al. (2002) use the 2QZ+6QZ data set to find
$\gamma$=0.42 and the constant $A=-11.10$. This normalisation is not
expected to be representative because their sample excludes resolved
objects, removing quasars with significant host galaxies. A range of
normalisations for the Croom et al. relation was determined using the
same quasar sample as above (Section \ref{PSFMag}). The resulting
relative frequency of quasars with a host galaxy contributing a given
fraction of the total brightness is shown in Fig. \ref{fig:Rgqhist}.
The same procedure was adopted to determine the range and relative
weightings of the constants $A$ for the third case, in which the
brightnesses of quasars and their host galaxies follow the Magorrian
relation, where $\gamma$ equals unity.


\section{Simulation Program}

All of the number-magnitude and number-redshift computations are
performed in IDL, with the equations for cosmological quantities such
as the distance modulus and comoving volume taken from Hogg
(2000). Each simulation in a particular band is minimally
parametrized by a QLF, bright and faint apparent magnitude limits, a
range in redshift, and the K-correction values as a function of
redshift. Additional options that may be specified include colour cuts
to duplicate realistic surveys, magnitude errors and a prescription
for the addition of host galaxy light to the quasars.

In real surveys there is generally some type of morphological
selection applied, aimed at eliminating objects that are dominated by
the host galaxy. To accommodate such selection, each simulation
includes a restriction on the maximum fraction of the total light from
the host galaxy, set at $R_{gq}<0.8$ (see Section \ref{PSFMag} for
definition). The $R_{gq}$ limit approximates the effect of
morphological selection. 

Every simulation also contains a faint absolute magnitude restriction
on the total brightness of each object, as listed in Table
\ref{tab:surveys}. When host galaxy light is added, an additional
faint magnitude limit of $M_{b_J}<-21$ is imposed on the quasar light
alone thus ensuring intrinsically faint AGN are not boosted into the
sample by the host galaxy light. 

Measurement errors on the magnitudes of the objects are also included.
The errors for $b_J$ are from Smith et al. (2005), and the SDSS DR3
catalogue (Schneider et al. 2005) for the SDSS $i$ and $K_{\rm 2MASS}$
passband errors. For the UKIDSS LAS, the errors are calculated using
the magnitudes quoted on the UKIDSS
website\footnote{http://www.ukidss.org/surveys/surveys.html}. Due to
the relatively small errors for most of the simulations ($\lesssim$10
per cent), the effect on the final results is small.

The specified redshift range is divided into slices, and at each slice
a luminosity function is created that covers the magnitudes accessible
at that redshift, thus providing the number of objects at each
magnitude at each redshift. If galaxy light is being added, the shape
of the luminosity function will change due to the extra flux from the
galaxies. The integral of the luminosity function at a specific
redshift slice gives the number of objects at that redshift, or
n(z). The sum over all redshifts within a small magnitude interval
produces the number of objects at each apparent magnitude, or n(m).


\section{Results}\label{Results}

As described above, a QLF, a representative quasar SED, a galaxy SED
and a relation between the magnitudes of the quasar and its host
galaxy have been combined to produce number-redshift and differential
number-magnitude counts for simulated surveys in the $b_{J}$, SDSS
$i$, WFCAM $Y$ and $K$, and $K_{\rm 2MASS}$ passbands, over the redshift
range $0.1<z<3.0$. Table \ref{tab:surveys} lists the parameters
adopted for each survey. The faint absolute magnitude limits applied
to every simulation are derived from the SDSS limit of $M_i<-22.4$,
which is then converted to the other bands via the $z=0$ quasar colours.

The following subsections outline the results of the simulations, and
highlight notable effects. As anticipated, host galaxy light has a
significant effect on number-redshift relations, particularly for
simulations in the near-infrared. The detailed shape of the galaxy SED has
little effect, whereas the results are very sensitive to the shape of
the quasar SED. Each prescription for adding host galaxy light
increases number counts, but the redshifts and magnitudes of the
objects most affected vary. Imposing a restriction on the fraction of
host galaxy compared to the total flux, which approximates a
morphological restriction, also becomes important, particularly at low
redshifts.

\begin{table*}
\centering
\caption{\label{tab:surveys}Simulated Surveys}
\begin{minipage}{18cm}
\begin{tabular}{cccccc} \hline
Survey Name & Magnitude Range & Magnitude Limit & Morphological Restriction &
 Quasar Only & Quasar + Galaxy\\
 & & & (input catalogue) & (N deg$^{-2}$) & (N deg$^{-2}$)\\ \hline

2QZ$+$6QZ & 16.0 $<b_J<$ 20.85 & M$_{b_J}<-22.0$ & Stellar & 52.4 & 52.9 \\
SDSS & 16.0 $<i<$ 20.5 & M$_i<-22.4$ & None & 63.7 & 66.1 \\
2MASS & 12.0 $<K_{\rm 2MASS}<$ 15.0 & M$_{K_{\rm 2MASS}}<-25.6$ & None & 0.28 & 0.52 \\
UKIDSS LAS$^{a}$ & 15.5 $<Y<$ 20.5 & M$_Y<-22.78$ & None & 81.7 & 88.1 \\
UKIDSS LAS & 14.0 $<K<$ 18.5 & M$_K<-25.7$ & None & 55.0 & 70.5 \\ \hline

\end{tabular}\\
$^{a}$ {\footnotesize Large Area Survey}\\
\end{minipage}
\end{table*}

\subsection{Initial Assessment}

\begin{figure}
  \resizebox{\hsize}{!}{\includegraphics{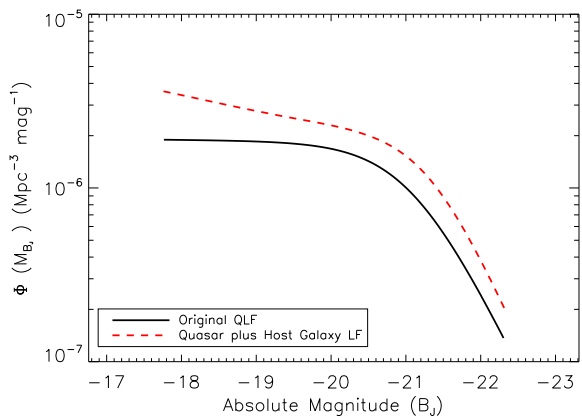}}
  \caption{QLF before and after adding light from the host galaxy at $z=0.1$.}
  \label{fig:LF}
\end{figure}

Fig. \ref{fig:LF} shows the effect that adding host galaxy light to the
flux of the quasars has on the shape of the QLF at $z=0.1$ and
relatively faint magnitudes. The slope of the faint end is steepened
significantly and the overall normalisation is increased. This effect
decreases with increasing redshift, as the quasars become brighter and
galaxy light is less significant.

\begin{figure}
  \resizebox{\hsize}{!}{\includegraphics{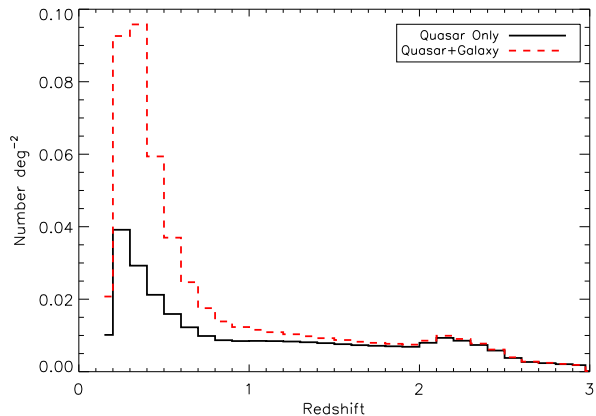}}
  \caption{Number-redshift, n(z), relationship for $12<K_{\rm 2MASS}<15$,
    $0.1<z<3.0$ simulation.}
  \label{fig:K2a}
\end{figure}

\begin{figure}
  \resizebox{\hsize}{!}{\includegraphics{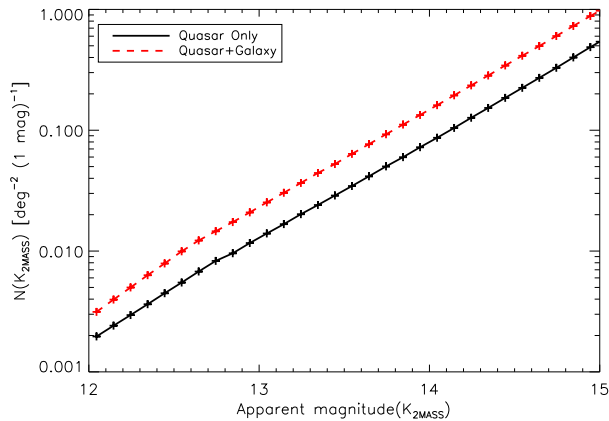}}
  \caption{Differential number-magnitude, n(m), relationship for
    $12<K_{\rm 2MASS}<15$, $0.1<z<3.0$ simulation.}
  \label{fig:K2b}
\end{figure}

\begin{figure}
  \resizebox{\hsize}{!}{\includegraphics{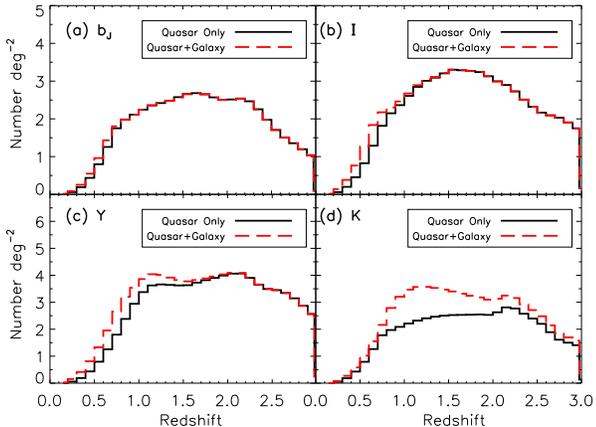}}
  \caption{Number-redshift, n(z), relationships for simulations listed
    in Table \ref{tab:surveys}. (a) 2QZ+6QZ $b_J$; (b) SDSS $i$; (c)
    UKIDSS LAS $Y$; (d) UKIDSS LAS $K$.}
  \label{fig:nzall}
\end{figure}

Figs. \ref{fig:K2a} and \ref{fig:K2b} are the number-redshift and
differential number-magnitude relationships, respectively, that result
from both the unaltered QLF and the LF when host galaxy light has been
added, for a relatively shallow simulation corresponding approximately
to 2MASS in the $K_{\rm 2MASS}$-band. The number-redshift counts are
significantly increased at low redshifts, with the high redshift
number counts remaining essentially the same. Fig. \ref{fig:K2b} shows
that the number of objects detected at every magnitude is
approximately doubled when host galaxy light is included. The
number-redshift relationships for the other simulations listed in
Table \ref{tab:surveys} are shown in Figs. \ref{fig:nzall}(a) through
(d), showing the increasing importance of host galaxy light at longer
wavelengths. The concave shape of the n(z) curves at redshifts less
than $\sim 1$ is due to the respective faint absolute magnitude
restrictions on each simulation. This cut is particularly significant
for the UKIDSS LAS $K$-band simulation, which has a relatively faint
apparent magnitude limit, and thus many low redshift objects are
eliminated. It is also worth noting in Fig. \ref{fig:nzall}(d), that
the presence of host galaxy light has a non-negligible effect on the
number counts at redshifts as high as $z\sim$2.

\subsection{Quasar SEDs}

\begin{figure}
  \resizebox{\hsize}{!}{\includegraphics{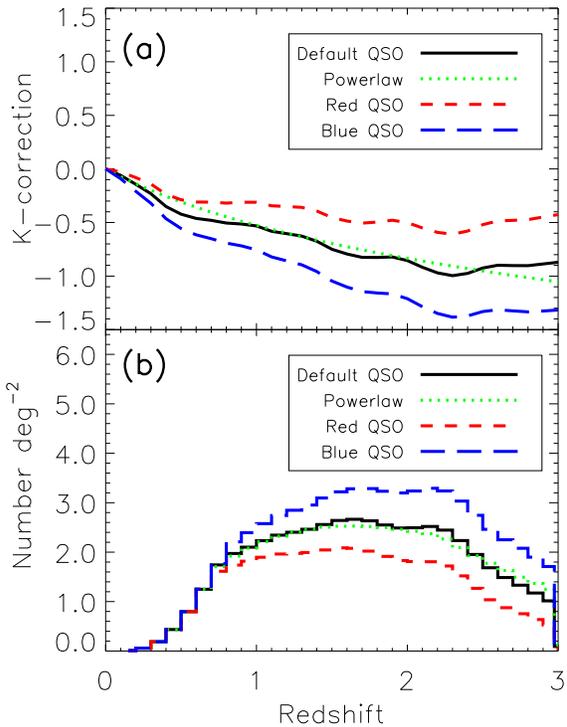}}
  \caption{Effect of quasar SED shape on number-redshift counts, for
    the 2QZ+6QZ simulation in $b_J$. (a)
    The default QSO SED, a pure power-law with $\alpha=-0.3$,
    red QSO SED with $\alpha=-0.6$, blue QSO SED with $\alpha=0.0$;
    (b) number-redshift counts for each of the SEDs.} 
  \label{fig:BjSED}
\end{figure}

\begin{figure}
  \resizebox{\hsize}{!}{\includegraphics{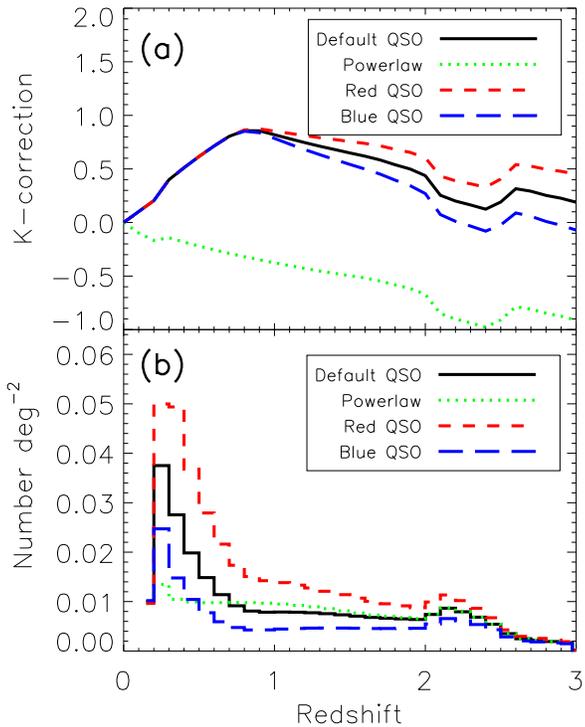}}
  \caption{Effect of quasar SED shape on number-redshift counts, for
    the 2MASS simulation in
    $K_{\rm 2MASS}$. (a) The default QSO SED, a power-law with
    $\alpha=-0.3$ and emission lines, red QSO SED with $\alpha=-0.6$,
    blue QSO SED with $\alpha=0.0$; (b) number-redshift counts for
    each of the SEDs.}
  \label{fig:K2SED}
\end{figure}

The importance of the overall SED shape is illustrated in Figs.
\ref{fig:BjSED} and \ref{fig:K2SED}. As seen in Fig.
\ref{fig:BjSED}(a), the K-correction derived from a pure power-law of
spectral index $\alpha=-0.3$ is a good approximation to the
K-correction from the default quasar SED in $b_J$ over the redshift
range $0<z<3$, except for the lack of features due to emission lines
within the passband. Fig. \ref{fig:BjSED}(b) shows that the bluer
($\alpha=0.0$) quasar SED increases the number counts at each redshift
and the redder ($\alpha=-0.6$) SED decreases the number counts, as
expected.

The situation is dramatically different for a near-infrared passband,
as shown for $K_{\rm 2MASS}$ in Fig. \ref{fig:K2SED}. Comparing
predictions for the default quasar SED against those for a quasar SED
where the blue $\alpha=-0.3$ power law continuum extends all the way
to $25\,000\,$\AA, illustrates the effect of the inflection in the
slope of the SED, at $\lambda \sim $12\,000\,\AA, on the low redshift
number counts. At $z>1$, the n(z)'s derived from the default and
power-law SEDs (Fig. \ref{fig:K2SED}(b), solid and dotted histograms)
come more into line, as the identical restframe optical portion of the
SEDs become increasingly important.

\subsection{Galaxy SEDs}

\begin{figure}
  \resizebox{\hsize}{!}{\includegraphics{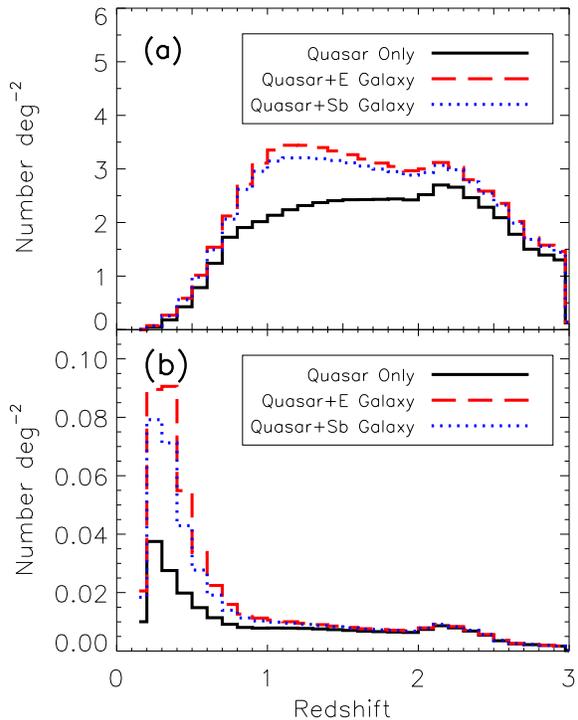}}
  \caption{Difference between adding an early type galaxy and an Sb
    galaxy to quasar flux. (a) Simulation corresponding to UKIDSS LAS in
    $K$; (b) Simulation corresponding to 2MASS in $K_{\rm 2MASS}$.
    Note the different y scales.}
  \label{fig:K2KUSb}
\end{figure}

Although it has been shown that using an appropriate quasar SED can
have significant effects on the number-redshift and number-magnitude
relations derived from the simulations, the exact form of the galaxy
SED incorporated has much less of an effect. Figs.
\ref{fig:K2KUSb}(a) and (b) show the relatively small impact in the
$K$-band, for both a deep and a shallow survey, due to the differences
between the elliptical and Sb host galaxy SEDs. The presence of host
galaxy light is much more important than the detailed shape of the SED.
The two curves converge at $z\approx 2$ due to the restframe
wavelength interval at which the quasar and host galaxy SEDs are
normalised moving into the $K$-band (see Section \ref{SED}).

\subsection{Quasar--Galaxy Relationships}\label{qgrelation2}

\begin{figure}
  \resizebox{\hsize}{!}{\includegraphics{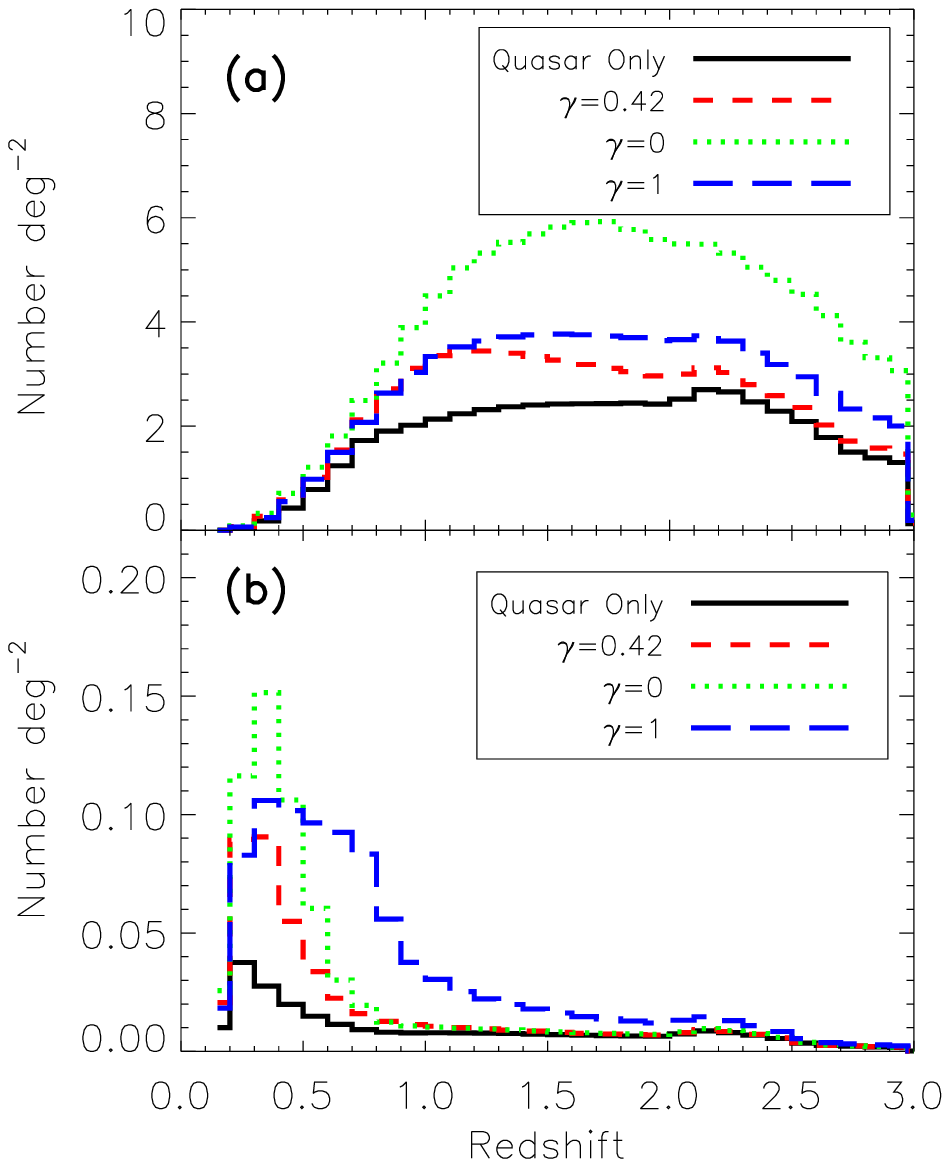}}
  \caption{Comparison of prescriptions for adding galaxy light. (a)
    Simulation corresponding to UKIDSS LAS in $K$; (b) Simulation
    corresponding to 2MASS in $K_{\rm 2MASS}$. Note the different y
    scales.}
  \label{fig:K2KUlumgals}  
\end{figure}

While the predictions are not very sensitive to the details of the
host galaxy SED, the prescription determining the brightness of the
host galaxy relative to the quasar has a major impact on the near-infrared
number counts (Fig. \ref{fig:K2KUlumgals}).

Setting $\gamma$=0, every quasar at a given redshift is assigned to
the same galaxy. This has very little effect when the galaxy is faint,
but when a relatively bright galaxy is added, the faint quasars are
now dominated by the host galaxy light. At higher redshifts, when all
of the objects are much brighter than the set galaxy magnitudes,
almost no effect is detectable, as seen in Fig.
\ref{fig:K2KUlumgals}(b).

The $\gamma$=1 case is very different. Each quasar at a given redshift
brightens in magnitude by the same amount when the galaxy flux is
added, effectively shifting the QLF brightward. Because neither faint
nor bright objects are preferentially affected in this scheme, the
number counts are increased almost equally at all redshifts, as seen
in Fig. \ref{fig:K2KUlumgals}(a).

Unsurprisingly, the $\gamma$=0.42 case produces intermediate
behaviour. Relatively faint quasars are assigned to relatively faint
host galaxies, while the brighter quasars end up in brighter galaxies,
with a smooth transition in between. The counts are increased at low
redshift, to a somewhat lesser degree than for $\gamma$=0, with a
decreasing effect as redshift increases and the quasars become much
brighter than their hosts.

\subsection{Galaxy Luminosity Evolution}\label{galbright}

In the case of the empirical correlations between host galaxy and
quasar luminosity, with $\gamma$=0.42 and $\gamma$=1, the BC2003 passive
evolution model adopted for the change in galaxy luminosity only affects
the maximum brightness a galaxy may obtain, at a given redshift.
None of the simulations are sensitive to sensible changes in the
adopted values for the maximum achievable galaxy luminosity.

\begin{figure}
  \resizebox{\hsize}{!}{\includegraphics{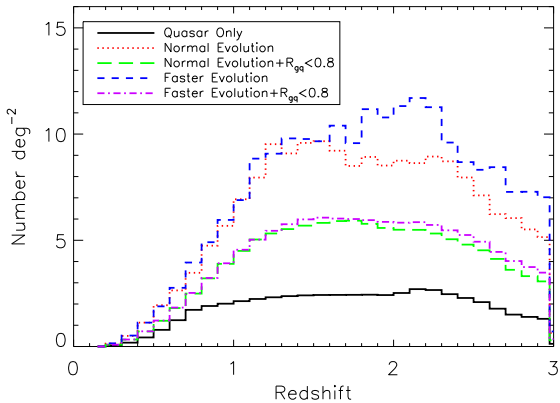}}
  \caption{Comparison of galaxy brightness evolution, with and without
  a morphological restriction $R_{gq}<0.8$ imposed. Simulation
  corresponding to UKIDSS LAS in $K$.}
  \label{fig:KUBC}
\end{figure}

In the case of the $\gamma$=0 model, with the galaxy luminosities
evolving according to the adopted BC2003 passive evolution
predictions, changes to the rate of evolution can have a significant
effect. As an illustration, the galaxies were allowed to become 0.5
mag brighter than the passive evolution case, modelled as a linear
increase in brightness as a function of lookback time over the range
$0.0 < z < 4.0$. There is little discernable effect at low redshifts,
as the luminosity increase is small, but for a near-infrared survey that probes
to faint quasar luminosities at significant redshifts, the
increase in numbers can be large. The origin of the increase is the
fraction of intrinsically faint quasars in luminous host galaxies,
where the increase in the host galaxy luminosity evolution brings
additional quasars above the faint magnitude limit (Fig.
\ref{fig:KUBC}). Such objects are, essentially by definition,
dominated by galaxy light and the inclusion of a morphological
restriction, such that $R_{gq}<0.8$, effectively removes the
additional population. The significant changes in the fraction of host
galaxy-dominated systems at high redshift in surveys to faint flux
limits in the near-infrared illustrates the potential importance of the host
galaxy on the number of quasars appearing above the survey flux limit.

\subsection{Morphological Selection Criteria}

It is clear from previous sections that the initial morphological
selection criteria applied to define a candidate quasar sample can be
important in certain circumstances. The conservative limit of
$R_{gq}<0.8$ adopted above is chosen to approximate the fraction of
host galaxy light that would result in an object being classified as a
galaxy rather than a quasar, and thus excluded from a quasar sample.
From Fig. \ref{fig:g-r}, it can be seen that for $R_{gq}>0.3$ almost
all objects are classified as extended in the SDSS catalogue.
Therefore, $R_{gq}=0.3$ may be taken as the approximate fraction of
galaxy light that is required for an object to appear non-stellar when
relatively high-quality imaging observations are available. 

\begin{figure}
  \resizebox{\hsize}{!}{\includegraphics{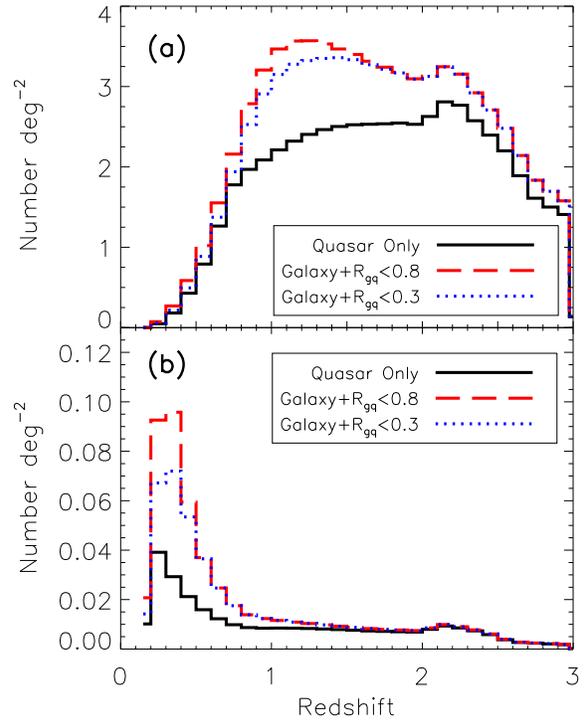}}
  \caption{Number-redshift relations for morphological selection
    criteria set at $R_{gq}<0.8$ and $R_{gq}<0.3$. (a) Simulation
    corresponding to UKIDSS LAS in $K$; (b) Simulation corresponding to
    2MASS in $K_{\rm 2MASS}$. Note the different y scales.}
  \label{fig:K2KUselect}
\end{figure}

Figs. \ref{fig:K2KUselect}(a) and (b) show the results from applying
the more stringent morphological restriction for the UKIDSS and 2MASS
$K$-band surveys, respectively, using the $\gamma$=0.42 host galaxy
prescription. The main difference appears at low redshifts, where the
quasars are of moderate luminosity and the host galaxies contribute
significantly to the total light. As anticipated, there is little
difference at high redshifts, where the objects are dominated by the
quasar light.

\section{Discussion}

\subsection{Initial Assessment}

Confidence in the results is gained by comparing the calculated
number-magnitude and number-redshift relations with results from
existing surveys. The excellent agreement between the predictions
incorporating the distribution of host galaxy luminosities from the
SDSS $i$-band sample, for 2QZ+6QZ in the $b_J$-band confirm that the
presence of host galaxies has not biased significantly the
determinations of the QLF from observations at blue wavelengths.

There is also good agreement between the predicted number counts in
the $i$-band and the published SDSS results to $i\sim 19.0$, which
indicates that the fraction of reddened quasars missing from a
$b_J$-selected sample but visible in the $i$-band, at nearly twice the
observed-frame wavelength, is not large.

As seen in Figs. \ref{fig:LF} through \ref{fig:nzall}, employing an
SED that represents the median quasar over a large range of
wavelengths and adding a prescribed amount of host galaxy light mostly
affect the low redshift ($z\lesssim 1$) number counts, particularly at
longer wavelengths. Hardly any effect is seen at $z\gtrsim 1.5$ except
for the UKIDSS LAS $K$-band simulation, where the faint magnitude
limits enable intrinsically faint quasars to be detected to high
redshifts. In fact, the UKIDSS LAS survey probes approximately one
magnitude fainter down the QLF than does the 2QZ survey.

Morphological and colour restrictions applied to define quasar
candidate samples need to be considered in order to avoid
contamination from the many low-redshift galaxy-dominated objects
while allowing the resolved quasars to be included. The results are
relevant to the design of future surveys as depths and areas required
to observe a given number of unobscured objects are now known. For
example, the area of sky in the UKIDSS LAS needed to establish the
existence of a given fraction of obscured quasars may be readily
calculated.

\subsection{Quasar SEDs}

The adoption of a single power-law approximation to the form of the
restframe quasar continuum over $1200 < \lambda < 5000\,$\AA \ has
proved largely adequate for surveys in the optical out to redshifts $z
\sim 2.5$. Extending observations into the near-infrared highlights the
inadequacy of the single power-law approximation. Incorporation of
the strong inflection in the SED at $\lambda \sim $12\,000\,\AA\ is
essential in order to understand observations at near-infrared wavelengths.
Fig. \ref{fig:K2SED}(a) illustrates the dramatic difference in the
K-correction at $z \lesssim 1$. The effect on the optical--NIR colours
of the quasars is large and the observed behaviour is sometimes
misattributed to the presence of the host galaxy making the colours
much redder, as in figure 7 of Glikman et al. (2004). Fig.
\ref{fig:SDSScolours} shows both the reddening of the optical--NIR
colour from the SED at $z \lesssim 1$ as well as the effect of the
host galaxy, seen only at $z \lesssim 0.3$. The shape of the SED and
the presence of host galaxy clearly produce different signatures.
Fig. \ref{fig:K2SED}(b) shows the combined impact on the number of
objects observed at $z\lesssim 1$ at bright near-infrared magnitudes.

The simulation predictions are very conservative in the sense that an
object contributing to the counts must have a minimum quasar absolute
magnitude of $M_{b_J}=-21.0$, equivalent to $M_{K_{\rm 2MASS}}=-24.6$.
The objects identified as AGN in 2MASS by Cutri et al. (2002) and
Francis et al. (2004) have no such limit applied. Furthermore, while
a $J-K > 2.0$ is often taken as a definition of a `red' AGN, our
default quasar SED has $J_{\rm 2MASS}-K_{\rm 2MASS}=2.5$ at $z=0.0$ and a
non-negligible fraction of optically-selected quasars possess $J-K \ge
2.0$ at low redshift. The impact of adopting fainter absolute
magnitude limits for `quasar' samples, plus the effect of any
additional colour selection criteria applied, need to be carefully
assessed before reliable conclusions concerning any excess of objects
over that predicted on the basis of optical samples can be reached.

\subsection{Quasar--Galaxy Relationships}

Given our lack of knowledge concerning the distribution of host galaxy
luminosities as a function of quasar luminosity the relations adopted
in the simulations deliberately span an extended range in possible
behaviour. One extreme, ($\gamma$=1), is motivated by theoretical
arguments from known relations between the mass of black holes and the
mass of their host bulges. The distribution of $R_{gq}$ then
approximates the range of accretion rates and efficiencies present in
the population.

The other extreme, ($\gamma$=0), where there is no relationship
between quasar and host galaxy luminosity, approximates a situation in
which any underlying relation between quasar and galaxy properties is
overwhelmed by variations in accretion rates and efficiency (relative
to the Eddington luminosity).

Intermediate between the two extremes, the $\gamma$=0.42 case is
observationally motivated (Croom et al. 2002). Such an observed
relationship is perhaps not unexpected on theoretical grounds but the
general applicability of the Croom et al. (2002) relationship remains
to be established. The quasars included in their study cover a
relatively modest range in redshift/luminosity and there must still be
concerns over the exclusion of objects with substantial galaxy
contributions. The fractional galaxy contribution was estimated from
fibre spectra and Croom et al.'s claim that the fraction of host
galaxy light that enters the fibre is effectively constant with
redshift relies on assumptions concerning the host galaxy sizes and
should be investigated quantitatively. However, we have chosen to
adopt the $\gamma$=0.42 case as the default for the simulations.

The key result of the simulations, in Section \ref{qgrelation2}, is
the significant impact of the presence of host galaxy light on the
surveys in the near-infrared. The n(z) results for both of the $K$-band
simulations are shown in Fig. \ref{fig:K2KUlumgals}. The three
prescriptions for adding host galaxy light affect different portions
of the QLF at a given redshift. When $\gamma$=1, the galaxies
increase in brightness at the same rate as the quasars in this case,
thus the n(z) relation is affected to higher redshifts. As can be seen
in Fig. \ref{fig:K2KUlumgals}(a) the n(z) at $z \sim 2$ differ
significantly among the host galaxy prescriptions, offering the
prospect of constraining the relation between quasar and host galaxy
luminosity at high redshifts.

The $\gamma$=0 prescription is unique in that some relatively
low-luminosity quasars have bright host galaxies. The faint quasars
are then dominated by the host galaxy light, producing a dramatic
steepening of the observed QLF at faint luminosities. The effect is
most evident at low redshifts, $z < 1$, as the fractional contribution
of host galaxy light decreases with increasing redshift, as the
luminosity of the quasars increases faster than the passively evolving
host galaxies. Whether such host galaxy-dominated objects feature
prominently in a near-infrared survey depends on the morphological
restriction imposed, since many very faint quasars are boosted above
the faint magnitude limit of the simulation due almost entirely to
light from the host galaxy. Requiring $R_{gq}<0.8$ ensures that host
galaxy-dominated objects are excluded. Interpreting the results of
surveys where host galaxy dominated objects are detected depends on
the ability to estimate accurately the luminosity of the central
source alone, so that relatively weak AGNs located in bright galaxies
are not classified as quasars (c.f. Section \ref{PSFMag}). As an
illustration of the importance of accurately separating quasar from
host galaxy light, the study of White et al. (2003) found several red
objects in an $i$-band selected sample. However, two of the four
objects with the largest derived $E(B-V)$ values from their table 1,
(J1011+5205, J1021+5114), were classified as non-stellar by the SDSS
photometry. As both objects are at $z\sim1$, they must be dominated by
their host galaxies in order to be resolved, and the red colours of
the objects almost certainly result in part from host galaxy light.

\subsection{Selection Criteria and Survey Design}\label{select}

Morphological and colour selection applied to an object catalogue
usually form an integral part of the identification of quasar
candidates. A morphological restriction on object selection is
approximated in the simulations by monitoring the fraction of host
galaxy flux that is contributing to the total flux of each object,
denoted $R_{gq}$. This restriction is only an approximation, as in
practice the $R_{gq}$ value separating morphological classifications
will be dependent on both the redshifts of the targets and the seeing
conditions of the observations. 

In the case of the 2QZ+6QZ survey, based on (by modern standards)
relatively poor quality photographic imaging, the adoption of strict
morphological selection turns out to have little impact on the
fraction of the quasar population included. Changing the morphological
restriction from $R_{gq}<0.8$ to $R_{gq}<0.3$ has almost no effect on
the counts. The insensitivity to the presence of host galaxies arises
due to a combination of the increasingly large contrast between quasar
and galaxy flux at ultraviolet wavelengths and the relatively high
lower redshift limit of $z_{low}=0.4$ imposed by Croom et al. (2002).

The same is not true for surveys in the near-infrared, as seen in Fig.
\ref{fig:K2KUselect}. Excluding extended sources will certainly
eliminate many low redshift quasars. The effect can also be seen by
using the 2MASS magnitudes for the objects in the SDSS DR3 quasar
catalogue. If the restriction of requiring point-like morphology is
imposed on the DR3 quasars, n(m) and n(z) counts in $K_{\rm 2MASS}$
agree well with the quasar-only $K_{\rm 2MASS}$ simulation, further
reinforcing the importance of adopting a realistic quasar SED
parametrization. Relaxing the point-like restriction, to include
resolved objects, results in n(m) and n(z) counts in $K_{\rm 2MASS}$
that agree well with the $K_{\rm 2MASS}$ simulation that includes host
galaxy light limited at $R_{gq}<0.8$. The study by Francis et al.
(2004), searching for dust-reddened quasars, selected objects from the
2MASS catalogue and found a higher surface density of $1\pm{0.3}$
quasar deg$^{-2}$ to $K_{\rm 2MASS}=15.0$. They note that some of
their `quasars' have optical colours indistinguishable from galaxies,
and the objects are dominated by their host galaxies. They impose no
morphological restriction on their sample, and some of their objects
could easily have $R_{gq}$ values approaching unity and would thus be
removed from our simulations.

It is clear from the discussion above that the inclusion of certain
non-stellar objects in quasar candidate samples selected in the
near-infrared will be necessary for many investigations. Of
particular interest is the impact of the presence of host galaxy light
on the colours of quasars and, thereby, on the effectiveness of
colour-based selection schemes designed to isolate particular elements
of the quasar population. For example, a key goal, now that faint
$K$-band data from UKIDSS LAS and other surveys are becoming
available, is to establish the fraction of reddened luminous quasars.

\begin{figure}
  \resizebox{\hsize}{!}{\includegraphics{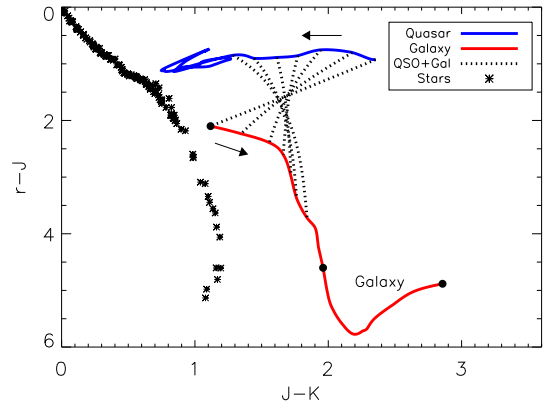}}
  \caption{Colour-colour diagram showing the evolution of the quasar
    locus (top) for $0.1<z<3.0$ and galaxy locus (bottom) for
    $0.1<z<2.0$, as well as the location of the stellar locus for
    dwarf stars from the Bruzual-Persson-Gunn-Stryker Atlas included
    in the STSDAS {\tt SYNPHOT} package. The dotted lines are for
    objects with $R_{gq}$ increasing from zero to unity, for redshifts
    0.1, 0.2, 0.3, 0.4, 0.5, 0.6, and 0.7. Arrows indicate the
    direction of increasing redshift for the quasar and galaxy
    tracks, with the filled circles on the galaxy track marking
    redshifts 0.1, 1.0 and 2.0.}
  \label{fig:ccgal}
\end{figure}

Warren, Hewett \& Foltz (2000) introduced the KX-method of selection
(by analogy with the UV-excess selection, or UVX) specifically to
address the problem of isolating reddened quasars using a combination
of optical and near-infrared colours. However, they did not consider
the impact of host galaxy light on the proposed selection scheme. Fig.
\ref{fig:ccgal} shows the colour-colour diagram advocated by Warren et
al. (2000). Shown are the default quasar and elliptical galaxy loci
for redshifts $0.1<z<3.0$ and $0.1<z<2.0$, respectively, the locus of
galactic dwarf stars, as well as quasar tracks with $0<R_{gq}<1$ at
several redshifts. The crucial aspect of the KX-method is that
reddened quasars move further away from the locus of galactic stars,
which will be present in enormous numbers relative to the much rarer
quasars. It can be seen immediately that the presence of the host
galaxies moves the quasars redward in $r-J$. Of more potential concern
is the horizontal displacement in $J-K$. At a fixed redshift, the
locus of objects with $0<R_{gq}<1$ is indicated by the dotted line,
joining the track (blue line) for the pure quasar, $R_{gq}=0$, to the
track (red) for the pure galaxy, $R_{gq}=1$.

At low redshifts, $z \lesssim 0.3$, increasing the amount of host
galaxy light makes the objects more blue in $J-K$, so very low
redshift, $z < 0.1$, galaxy-dominated objects will be excluded with a
moderate colour cut, like the $J-K>1.2$ selection employed by Francis
et al. (2004). Fortunately, the effect of the host galaxy light
diminishes rapidly with increasing redshift and only for redshifts
$z<0.1$ is proximity to the stellar locus an issue for the quasar
selection. At $z\gtrsim 0.3$, increasing host galaxy contribution
causes the objects to appear redder in $J-K$, improving the
effectiveness of the KX-selection scheme. Note however that the sense
of the colour changes for quasars with $0.5 \lesssim z \lesssim 2$ are
very similar to those due to simply reddening a pure quasar SED,
highlighting the importance of quantifying the fraction of host galaxy
light present when interpreting surveys for reddened quasars.

\section{Summary}

Simulations of quasar surveys ranging from optical to near-infrared
wavelengths have been performed by combining current information on
the quasar luminosity function and the quasar SED. Host galaxy light
was added according to different prescriptions and has the most effect
at longer wavelengths. The main results of this study are:
\begin{itemize}
\item The overall shape of the quasar SED is important, particularly
  the correct treatment of the continuum longward of
  12\,000\,\AA\, for the surveys at near-infrared wavelengths. 
\item The near-infrared simulations produced many more
  galaxy-dominated objects than simulations at optical wavelengths and
  were much more affected by morphological restrictions, which should
  thus be chosen carefully if performing surveys at longer wavelengths. 
\item The simulations are not sensitive to the SED of the galaxy added to
  the quasar, but are sensitive to the prescription relating host
  galaxy and quasar luminosity. 
\end{itemize}
The predictions presented should provide a reference for the
contribution of the unobscured quasar population in near-infrared
surveys, and are of direct relevance to the design of future surveys.

\section*{Acknowledgments}

This publication makes use of data products from the Two Micron All
Sky Survey, which is a joint project of the University of
Massachusetts and the Infrared Processing and Analysis
Center/California Institute of Technology, funded by the National
Aeronautics and Space Administration and the National Science
Foundation.

Funding for the Sloan Digital Sky Survey (SDSS) has been provided by
the Alfred P. Sloan Foundation, the Participating Institutions, the
National Aeronautics and Space Administration, the National Science
Foundation, the U.S. Department of Energy, the Japanese
Monbukagakusho, and the Max Planck Society. The SDSS Web site is
http://www.sdss.org/.

The SDSS is managed by the Astrophysical Research Consortium (ARC) for
the Participating Institutions. The Participating Institutions are The
University of Chicago, Fermilab, the Institute for Advanced Study, the
Japan Participation Group, The Johns Hopkins University, Los Alamos
National Laboratory, the Max-Planck-Institute for Astronomy (MPIA),
the Max-Planck-Institute for Astrophysics (MPA), New Mexico State
University, University of Pittsburgh, Princeton University, the United
States Naval Observatory, and the University of Washington.

We thank the referee, Paul Francis, for improving the presentation of
this paper, and Eilat Glikman for providing the optical--NIR quasar
composite spectrum in advance of publication. NM wishes to thank the
Overseas Research Students Awards Scheme, the Cambridge Commonwealth
Trust, and the Dr. John Taylor Scholarship from Corpus Christi College
for their generous support.

\clearpage

\appendix

\section{The b$_J$ passband}

The $b_J$ passband was key in the compilation of the second period of
all-sky photographic surveys that began in the 1970s with the
programmes using the 1.2m United Kingdom Schmidt Telescope at Siding
Spring in Australia (Cannon 1984) and continued with the initiation
of the Palomar Observatory Sky Survey II (Reid et al. 1991) using the
1.2m Oschin Schmidt at Palomar Mountain in the United States of
America. The resulting all-sky surveys have formed a critical part of
modern astrophysical research, particularly in the form of the
products from the digital scanning of the photographic plates (Lasker
1995).

Conventional colour equations relating $b_J$ to Johnson $B$ and $V$
were established at an early stage (Blair \& Gilmore 1982). However,
to our knowledge, no published determination of the response function
for the $b_J$ passband exists in the astronomical literature. One of
us (Hewett) made such a determination more than a decade ago and the
resulting response function has been used in many investigations,
frequently with a reference to `Hewett (private
communication)'. 

The basis for the determination of the $b_J$ passband is the
unfiltered objective-prism plate UJ10738P, obtained on 1986 February
11, at a mean airmass of 1.3, with the United Kingdom Schmidt
Telescope. The 55 minute exposure, on sensitised Kodak IIIaJ emulsion,
of a portion of the Virgo Cluster was scanned by the APM Facility as
part of the Large Bright Quasar Survey (Hewett, Foltz \& Chaffee
1995). The follow-up observations of candidate quasars undertaken at
the Multiple Mirror Telescope resulted in a large number of fluxed
spectra of quasars and stars covering the wavelength range from the
atmospheric cutoff at $\sim3200\,$\AA \ to $\sim 7500\,$\AA, well
beyond the red cutoff of the Kodak IIIaJ emulsion. It was therefore
possible to make a direct empirical determination of the IIIaJ
emulsion sensitivity by comparing the fluxed object spectra with the
individual intensity scans of the corresponding objective-prism
spectra from the APM.

The passband presented here is based on the use of $\sim120$ spectra,
each with corresponding unsaturated objective-prism spectra from the
APM scans. The fluxed spectra were multiplied by an approximation to
the wavelength dependent passband transmission (taken to be constant
initially). A cutoff to the emulsion response, linear in wavelength,
at the red end of the spectra was incorporated. The spectra were then
rebinned to match the non-linear wavelength scale appropriate to the
objective-prism spectra and convolved with a Moffat profile (Moffat
1969), to take account of the seeing. The resulting spectra were then
compared to the corresponding objective-prism spectra. The fluxed and
objective-prism spectra were normalised using the wavelength interval
$4000-5000\,$\AA. An approximation to the transmission at each
wavelength was made by taking the median value at each wavelength of
the 120 factors obtained by dividing each object-prism spectrum by the
synthetic spectrum.

A $\chi^2$ measure of the goodness-of-fit for the new approximation to
the transmission as a function of wavelength could then be obtained by
including the new estimate at the initial stage of the process and
calculating a $\chi^2$ value, over the full wavelength range for which
the transmission is non-zero, using the resulting synthetic and actual
objective-prism spectra. The procedure was incorporated into the {\tt
AMOEBA} routine from Numerical Recipes (Press et al. 1992) to
determine the best-fit parameters describing the wavelength cutoff
and the seeing, along with the wavelength dependent transmission.
Technically, the wavelength dependent transmission should be
determined independently by the routine, prior to blurring by the
seeing. However, in practice, away from the relatively sharp emulsion
cutoff, the effect of the seeing on the slowly varying transmission
function is simply to apply a modest degree of smoothing. The
synthetic spectra produced employing the final transmission function
provide a superb match to the individual objective-prism spectra.

\begin{figure}
  \resizebox{\hsize}{!}{\includegraphics{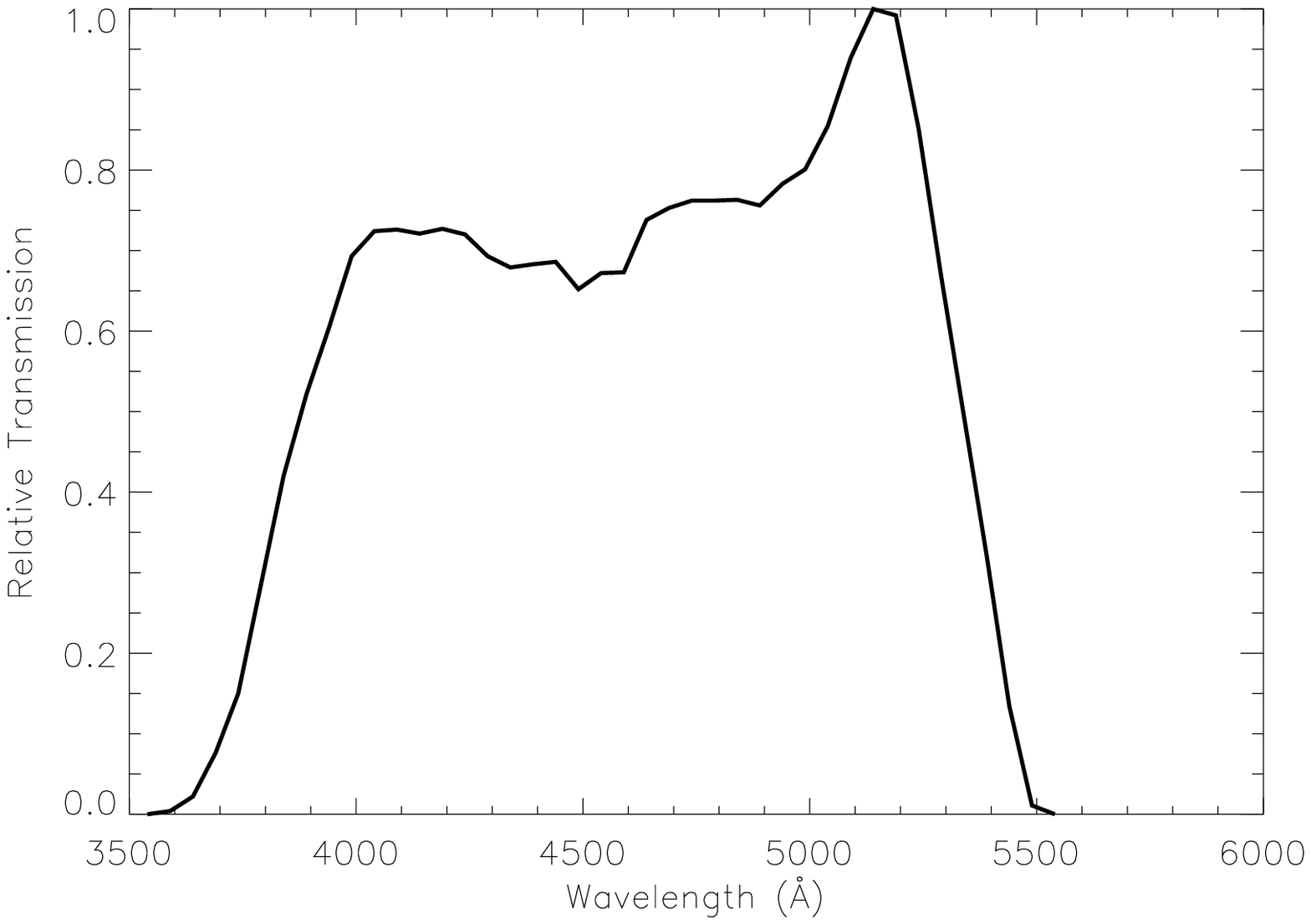}}
  \caption{$b_J$ relative transmission function for energy integration
    detectors.}
  \label{fig:filter}
\end{figure}

The resulting transmission function (Figure \ref{fig:filter})
represents the combination of atmosphere, telescope optics and the
photographic emulsion. Transforming to a function appropriate to
unit-, or zero-, airmass was achieved by correcting for extinction
using the mean extinction curve for Kitt Peak, which is expected to be
a good approximation for the Siding Spring site.

The objective-prism plate was obtained without any short-wavelength
filter, with the decreasing transmission towards the blue due mainly
to the effect of atmospheric extinction. The appropriate blue cutoff
for the $b_J$ passband was applied using the transmission curve for a
2mm thick Schott GG395 filter. Note that the similar `$b_J$'
passband employed at several 4m-class telescopes in the 1980s
employed a GG385 filter (e.g. Kron 1980).

The resulting transmission function is appropriate for calculating the
magnitudes of objects recorded using detectors, such as photographic
plates, that integrate energy per unit wavelength across the
passband. The majority of modern detectors, such as CCDs, integrate
photon counts across the passband and many synthetic photometry
packages, such as STSDAS {\tt SYNPHOT}, assume by default that the
detector is a photon-counting device (see Bessell, Castelli \& Plez
(1998) for a more detailed discussion). In order to obtain the correct
results for the $b_J$ passband using such packages it is necessary to
modify the input transmission function, scaling the transmission by a
factor of $1/\lambda$. Table \ref{tab:trans} contains tabulations of the
normalised $b_J$ passband transmissions suitable for use with energy-
and photon-integration software packages.

\begin{table}
\caption{\label{tab:trans}Relative Transmission for the $b_J$ Passband}
\begin{tabular}{ccc} \hline
Wavelength & Relative Transmission & Relative Transmission$^a$ \\ 
(\AA) & &  \\ \hline
 3540 & 0.000 & 0.000 \\
 3590 & 0.004 & 0.006 \\
 3640 & 0.022 & 0.031 \\
 3690 & 0.076 & 0.105 \\
 3740 & 0.150 & 0.207 \\
 3790 & 0.285 & 0.386 \\
 3840 & 0.420 & 0.562 \\
 3890 & 0.521 & 0.688 \\
 3940 & 0.604 & 0.788 \\
 3990 & 0.693 & 0.893 \\
 4040 & 0.724 & 0.920 \\
 4090 & 0.726 & 0.912 \\
 4140 & 0.721 & 0.896 \\
 4190 & 0.727 & 0.891 \\
 4240 & 0.720 & 0.873 \\
 4290 & 0.693 & 0.830 \\
 4340 & 0.679 & 0.804 \\
 4390 & 0.683 & 0.800 \\
 4440 & 0.686 & 0.794 \\
 4490 & 0.652 & 0.746 \\
 4540 & 0.672 & 0.761 \\
 4590 & 0.673 & 0.754 \\
 4640 & 0.738 & 0.817 \\
 4690 & 0.753 & 0.825 \\
 4740 & 0.762 & 0.826 \\
 4790 & 0.762 & 0.817 \\
 4840 & 0.763 & 0.810 \\
 4890 & 0.756 & 0.794 \\
 4940 & 0.783 & 0.815 \\
 4990 & 0.801 & 0.825 \\
 5040 & 0.855 & 0.872 \\
 5090 & 0.939 & 0.949 \\
 5140 & 1.000 & 1.000 \\
 5190 & 0.992 & 0.982 \\
 5240 & 0.851 & 0.834 \\
 5290 & 0.667 & 0.648 \\
 5340 & 0.494 & 0.476 \\
 5390 & 0.321 & 0.306 \\
 5440 & 0.134 & 0.127 \\
 5490 & 0.011 & 0.010 \\
 5540 & 0.000 & 0.000 \\
\hline
\end{tabular}
$^{a}$ {\footnotesize Transmission values for use with software that assumes
a photon counting detector}\\
\end{table}

\clearpage

\section{Differential Number-Magnitude Tables}

Tables containing the differential number-magnitude counts for each of
the surveys simulated are given in Tables \ref{tab:Bjresults} through
\ref{tab:KUresults}. The magnitude and redshift limits of the
simulations are given in the caption of each table, and may also be
found in Section \ref{Results}. The definition of $R_{gq}$ is
described in Section \ref{PSFMag}, and the prescription for adding
host galaxy light to that of the quasar is taken from Croom et
al. (2002), with $\gamma$=0.42. In each table, the first column gives the
magnitude range for each entry, and the second column lists the number
of objects predicted by simulations including only the quasar SED
between the listed magnitudes. The third and fourth columns list the
number of objects predicted by simulations that include host galaxy
light, with limits of 80 and 30 per cent, respectively, imposed on the final
galaxy light contributions.


\begin{table}
\centering
\caption{\label{tab:Bjresults}Differential number per 0.25 magnitude,
  deg$^{-2}$ for the survey in $b_J$, with magnitude limits of
  $16.0<b_J<20.85$ and a faint absolute magnitude limit for the total
  combined flux from quasar and host galaxy of $M_{b_J}<-22.0$, and
  redshifts $0.1<z<3.0$.}
\begin{tabular}{cccc} \hline
Magnitude & Quasar Only & Q+G & Q+G \\ 
 & & $R_{gq}<0.8$ & $R_{gq}<0.3$ \\ \hline
16.00 - 16.24 &  0.010 &  0.010 &  0.010 \\
16.25 - 16.49 &  0.016 &  0.017 &  0.017 \\
16.50 - 16.74 &  0.027 &  0.027 &  0.027 \\
16.75 - 16.99 &  0.045 &  0.046 &  0.045 \\
17.00 - 17.24 &  0.073 &  0.074 &  0.074 \\
17.25 - 17.49 &  0.121 &  0.123 &  0.122 \\
17.50 - 17.74 &  0.195 &  0.198 &  0.196 \\
17.75 - 17.99 &  0.315 &  0.319 &  0.317 \\
18.00 - 18.24 &  0.514 &  0.519 &  0.515 \\
18.25 - 18.49 &  0.812 &  0.819 &  0.813 \\
18.50 - 18.74 &  1.233 &  1.243 &  1.235 \\
18.75 - 18.99 &  1.887 &  1.900 &  1.887 \\
19.00 - 19.24 &  2.730 &  2.749 &  2.731 \\
19.25 - 19.49 &  3.760 &  3.787 &  3.761 \\
19.50 - 19.74 &  4.986 &  5.023 &  4.987 \\
19.75 - 19.99 &  6.254 &  6.305 &  6.257 \\
20.00 - 20.24 &  7.514 &  7.581 &  7.517 \\
20.25 - 20.49 &  8.418 &  8.512 &  8.430 \\
20.50 - 20.74 &  9.199 &  9.318 &  9.214 \\
20.75 - 20.85 &  4.291 &  4.354 &  4.300 \\ \hline
Total & 52.40 & 52.93 & 52.46 \\
\hline
\end{tabular}\\
\end{table}

\begin{table}
\centering
\caption{\label{tab:Iresults}Differential number per 0.25 magnitude,
  deg$^{-2}$ for the survey in $i$, with magnitude limits of
  $16.0<i<20.5$ and a faint absolute magnitude limit for the total
  combined flux from quasar and host galaxy of $M_{i}<-22.4$, and
  redshifts $0.1<z<3.0$.} 
\begin{tabular}{cccc} \hline
Magnitude & Quasar Only & Q+G & Q+G \\ 
 & & $R_{gq}<0.8$ & $R_{gq}<0.3$ \\ \hline
16.00 - 16.24 &  0.033 &  0.039 &  0.037 \\
16.25 - 16.49 &  0.053 &  0.063 &  0.060 \\
16.50 - 16.74 &  0.087 &  0.101 &  0.097 \\
16.75 - 16.99 &  0.143 &  0.164 &  0.158 \\
17.00 - 17.24 &  0.234 &  0.264 &  0.256 \\
17.25 - 17.49 &  0.380 &  0.425 &  0.413 \\
17.50 - 17.74 &  0.612 &  0.675 &  0.659 \\
17.75 - 17.99 &  0.971 &  1.057 &  1.033 \\
18.00 - 18.24 &  1.502 &  1.614 &  1.580 \\
18.25 - 18.49 &  2.250 &  2.385 &  2.338 \\
18.50 - 18.74 &  3.218 &  3.375 &  3.316 \\
18.75 - 18.99 &  4.380 &  4.557 &  4.482 \\
19.00 - 19.24 &  5.640 &  5.837 &  5.744 \\
19.25 - 19.49 &  6.888 &  7.109 &  6.999 \\
19.50 - 19.74 &  8.032 &  8.270 &  8.140 \\
19.75 - 19.99 &  8.984 &  9.234 &  9.079 \\
20.00 - 20.24 &  9.720 &  9.981 &  9.795 \\
20.25 - 20.50 & 10.591 & 10.908 & 10.669 \\ \hline
Total & 63.720 & 66.058 & 64.852 \\
\hline
\end{tabular}\\
\end{table}

\begin{table}
\centering
\caption{\label{tab:K2results}Differential number per 0.25 magnitude,
  deg$^{-2}$ for the survey in $K_{\rm 2MASS}$, with magnitude limits of
  $12.0<K_{\rm 2MASS}<15.0$ and a faint absolute magnitude limit for the total
  combined flux from quasar and host galaxy of $M_{K_{\rm 2MASS}}<-25.6$,
  and redshifts $0.1<z<3.0$.} 
\begin{tabular}{cccc} \hline
Magnitude & Quasar Only & Q+G & Q+G \\ 
 & & $R_{gq}<0.8$ & $R_{gq}<0.3$ \\ \hline
12.00 - 12.24 &  0.001 &  0.001 &  0.001 \\
12.25 - 12.49 &  0.001 &  0.002 &  0.001 \\
12.50 - 12.74 &  0.002 &  0.003 &  0.002 \\
12.75 - 12.99 &  0.003 &  0.005 &  0.004 \\
13.00 - 13.24 &  0.004 &  0.007 &  0.006 \\
13.25 - 13.49 &  0.006 &  0.012 &  0.010 \\
13.50 - 13.74 &  0.010 &  0.018 &  0.016 \\
13.75 - 13.99 &  0.016 &  0.029 &  0.025 \\
14.00 - 14.24 &  0.025 &  0.047 &  0.040 \\
14.25 - 14.49 &  0.040 &  0.075 &  0.065 \\
14.50 - 14.74 &  0.065 &  0.120 &  0.106 \\
14.75 - 15.00 &  0.111 &  0.203 &  0.182 \\ \hline
Total & 0.284 & 0.521 & 0.458 \\
\hline
\end{tabular}\\
\end{table}

\begin{table}
\centering
\caption{\label{tab:Yresults}Differential number per 0.25 magnitude,
  deg$^{-2}$ for the survey in $Y$, with magnitude limits of
  $15.5<Y<20.5$ and a faint absolute magnitude limit for the total
  combined flux from quasar and host galaxy of $M_{Y}<-22.78$, and
  redshifts $0.1<z<3.0$.} 
\begin{tabular}{cccc} \hline
Magnitude & Quasar Only & Q+G & Q+G \\ 
 & & $R_{gq}<0.8$ & $R_{gq}<0.3$ \\ \hline
15.50 - 15.74 &  0.024 &  0.033 &  0.030 \\
15.75 - 15.99 &  0.040 &  0.054 &  0.050 \\
16.00 - 16.24 &  0.067 &  0.088 &  0.083 \\
16.25 - 16.49 &  0.111 &  0.145 &  0.136 \\
16.50 - 16.74 &  0.182 &  0.236 &  0.224 \\
16.75 - 16.99 &  0.298 &  0.380 &  0.363 \\
17.00 - 17.24 &  0.483 &  0.605 &  0.579 \\
17.25 - 17.49 &  0.770 &  0.945 &  0.908 \\
17.50 - 17.74 &  1.204 &  1.440 &  1.389 \\
17.75 - 17.99 &  1.834 &  2.138 &  2.071 \\
18.00 - 18.24 &  2.695 &  3.062 &  2.977 \\
18.25 - 18.49 &  3.793 &  4.212 &  4.105 \\
18.50 - 18.74 &  5.045 &  5.526 &  5.399 \\
18.75 - 18.99 &  6.268 &  6.882 &  6.734 \\
19.00 - 19.24 &  7.544 &  8.128 &  7.956 \\
19.25 - 19.49 &  8.710 &  9.246 &  9.041 \\
19.50 - 19.74 &  9.690 & 10.207 &  9.946 \\
19.75 - 19.99 & 10.503 & 11.020 & 10.687 \\
20.00 - 20.24 & 11.010 & 11.608 & 11.219 \\
20.25 - 20.50 & 11.388 & 12.123 & 11.658 \\ \hline
Total & 81.658 & 88.076 & 85.555 \\
\hline
\end{tabular}\\
\end{table}

\begin{table}
\centering
\caption{\label{tab:KUresults}Differential number per 0.25 magnitude,
  deg$^{-2}$ for the survey in $K$, with magnitude limits of
  $14.0<K<18.5$ and a faint absolute magnitude limit for the total
  combined flux from quasar and host galaxy of $M_{K}<-25.7$, and
  redshifts $0.1<z<3.0$.} 
\begin{tabular}{cccc} \hline
Magnitude & Quasar Only & Q+G & Q+G \\ 
 & & $R_{gq}<0.8$ & $R_{gq}<0.3$ \\ \hline
14.00 - 14.24 &  0.027 &  0.050 &  0.043 \\
14.25 - 14.49 &  0.043 &  0.078 &  0.069 \\
14.50 - 14.74 &  0.067 &  0.122 &  0.109 \\
14.75 - 14.99 &  0.106 &  0.191 &  0.172 \\
15.00 - 15.24 &  0.167 &  0.296 &  0.270 \\
15.25 - 15.49 &  0.262 &  0.455 &  0.420 \\
15.50 - 15.74 &  0.410 &  0.695 &  0.649 \\
15.75 - 15.99 &  0.636 &  1.054 &  0.993 \\
16.00 - 16.24 &  0.977 &  1.571 &  1.493 \\
16.25 - 16.49 &  1.477 &  2.297 &  2.198 \\
16.50 - 16.74 &  2.182 &  3.269 &  3.146 \\
16.75 - 16.99 &  3.128 &  4.488 &  4.335 \\
17.00 - 17.24 &  4.305 &  5.900 &  5.715 \\
17.25 - 17.49 &  5.651 &  7.405 &  7.181 \\
17.50 - 17.74 &  7.077 &  8.903 &  8.630 \\
17.75 - 17.99 &  8.474 & 10.300 &  9.970 \\
18.00 - 18.24 &  9.575 & 11.332 & 10.939 \\
18.25 - 18.50 & 10.484 & 12.133 & 11.665 \\ \hline
Total & 55.049 & 70.537 & 67.998 \\
\hline
\end{tabular}\\
\end{table}


\end{document}